\documentclass[usenatbib]{mnras}
\usepackage{newtxtext,newtxmath}

\usepackage[T1]{fontenc}
\usepackage{ae,aecompl}
\usepackage{macros}

\usepackage{geometry}
\usepackage{fancyhdr}
\usepackage{amsmath,amssymb}
\usepackage{graphicx} 
\usepackage[caption=false]{subfig}
\usepackage{hyperref}
\usepackage{wasysym}
\usepackage{lastpage}
\usepackage{etoolbox}
\usepackage{threeparttable}
\usepackage{lscape}
\usepackage{multicol}
\usepackage{academicons}

\graphicspath{{./}{Figures/}}

\title[Testing orbit modeling of satellites]{Modeling the orbital histories of satellites of Milky Way-mass galaxies: testing static host potentials against cosmological simulations}

\author[Santistevan et al.]{
Isaiah B. Santistevan,$^{1}$\thanks{E-mail: ibsantistevan@ucdavis.edu},
Andrew Wetzel$^{1}$,
Erik Tollerud$^{2}$,
Robyn E. Sanderson$^{3,4}$,
Jorge Moreno$^{4,5}$,
\newauthor
Ekta Patel$^6$\thanks{Hubble Fellow}\\
\\
$^{1}$Department of Physics \& Astronomy, University of California, Davis, CA 95616, USA\\
$^{2}$Space Telescope Science Institute, 3700 San Martin Dr, Baltimore, MD 21218, USA\\
$^{3}$Department of Physics \& Astronomy, University of Pennsylvania, Philadelphia, PA 19104, USA\\
$^{4}$Center for Computational Astrophysics, Flatiron Institute, New York, NY 10010, USA\\
$^{5}$Department of Physics and Astronomy, Pomona College, Claremont, CA 91711, USA\\
$^{6}$Department of Physics and Astronomy, University of Utah, 115 South 1400 East, Salt Lake City, Utah 84112, USA\\
}

\date{Accepted XXX. Received YYY; in original form ZZZ}
\pubyear{2023}

\begin{document}
\label{firstpage}
\pagerange{\pageref{firstpage}--\pageref{lastpage}}
\maketitle

\begin{abstract}
Understanding the evolution of satellite galaxies of the Milky Way (MW) and M31 requires modeling their orbital histories across cosmic time.
Many works that model satellite orbits incorrectly assume or approximate that the host halo gravitational potential is fixed in time and is spherically symmetric or axisymmetric.
We rigorously benchmark the accuracy of such models against the FIRE-2 cosmological baryonic simulations of MW/M31-mass halos.
When a typical surviving satellite fell in ($3.4-9.7\Gyr$ ago), the host halo mass and radius were typically $26-86$ per cent of their values today, respectively.
Most of this mass growth of the host occurred at small distances, $r\lesssim50\kpc$, opposite to dark-matter-only simulations, which experience almost no growth at small radii.
We fit a near-exact axisymmetric gravitational potential to each host at $z=0$ and backward integrate the orbits of satellites in this static potential, comparing against the true orbit histories in the simulations.
Orbital energy and angular momentum are not well conserved throughout an orbital history, varying by 25 per cent from their current values already $1.6-4.7\Gyr$ ago.
Most orbital properties are minimally biased, $\lesssim10$ per cent, when averaged across the satellite population as a whole.
However, for a single satellite, the uncertainties are large: recent orbital properties, like the most recent pericentre distance, typically are $\approx20$ per cent uncertain, while earlier events, like the minimum pericentre or the infall time, are $\approx40-80$ per cent uncertain.
Furthermore, these biases and uncertainties are lower limits, given that we use near-exact host mass profiles at $z=0$.
\end{abstract}




\section{Introduction} %
\label{sec:intro}      %

The satellite galaxies of the Milky Way (MW) and M31 are the most rigorously studied low-mass galaxies, given their proximity to us.
The dynamics and evolution of these low-mass galaxies encode rich information about their past and the host halo environment in which they orbit.
These low-mass galaxies also differ from the non-satellite galaxies within the Local Group (LG) given that their host galaxies, either the MW or M31, regulated their star formation, and they probe deep within their host potentials.
Important questions about their orbital histories include:
When did each satellite fall into the MW/M31 halo?
How close have they orbited, and when did they experience pericentric passages?
How has the mass of the MW/M31 changed over time, and how has it impacted satellite orbits?
How well conserved are orbital properties such as energy or angular momentum?
Given a near-perfect representation of the host potential, to what extent can we accurately recover orbit properties, such pericentre distances and times?

Satellites in the Local Group (LG) are the only low-mass galaxies for which we can measure their full 6D phase-space coordinates.
One historically challenging phase-space component to measure is proper motion.
Many studies used the Hubble Space Telescope (HST) to derive proper motion estimates for galaxies in the LG to estimate their star-formation histories, determine companionship with the Large Magellanic Cloud (LMC), and to study the planar structure of satellites around both the MW and Andromeda (M31) \citep[for example][]{vanderMarel12, Kallivayalil13, Sohn20, Pawlowski21}.
Recent HST treasury programs (such as GO-14734, PI Kallivayalil; GO-15902, PI Weisz; GO-17174, Bennet; HSTPROMO) are obtaining proper motions for the remaining satellites; we soon will have orbital dynamics information for all known satellites in the LG.

Furthermore, the \textit{Gaia} space telescope has been revolutionary in providing a wealth of data, such as positions, magnitudes, and proper motions, for over 1 billion sources, including globular clusters and the satellite galaxies of the MW \citep{GaiaDR2}.
Numerous studies use Gaia's kinematic data to study group infall of satellites, including satellites of the LMC \citep[for example][]{Kallivayalil18, Fritz18, Patel20}.
Because proper motion measurements improve with multiple observations, many studies now use HST and Gaia data in conjunction to reduce uncertainties in satellite galaxy proper motions \citep[such as][]{delPino22, Bennet22, Warfield23}.
Current observational programs such as the Satellites Around Galactic Analogs (SAGA) survey \citep{SAGA_I, SAGA_II} are observing satellite galaxies around other MW-mass galaxies.
The James Webb Space Telescope (JWST) soon will be able to obtain proper motions of even more distant low-mass galaxies, beyond the LG \citep[][]{Weisz23}, and the Vera Rubin Telescope will catalog more than 10 billion stars within the low-mass galaxies around the MW.

Using the phase-space information of satellite galaxies, in tandem with a model of the Galactic potential, many studies investigate satellite infall and orbital histories.
A galaxy becomes a satellite galaxy when it first crosses the virial radius of a more massive halo, which can quench the lower-mass galaxy's star formation \citep[for example][]{Gunn72, vandenBosch08, RodriguezWimberly19, Samuel22, Samuel22_ram}.
As satellites reach their closest approach to the host galaxy at pericentre, they orbit in the denser host CGM and feel strong tidal forces ram pressure \citep[for example][]{McCarthy08, Simons20, MartinNavarro21, Samuel22_ram}.
Many studies use different models for the MW/M31 potential and numerically integrate the orbits for their satellite galaxies to derive orbit properties. However, the results depend strongly on modeling the total mass profile of the MW  and M31 \citep[for example][]{Gaia_orbits, Fritz18, Fritz18_Segue}.
Another study by \cite{Fillingham19} jointly used Gaia data with the Phat ELVIS \citep[pELVIS;][]{Kelley19} dark matter only (DMO) simulations, which include the gravitational effects of a central galaxy, to match simulated satellites to observed satellites in 6D phase space.
They then used the distribution of infall times of the matched simulated satellites to infer the infall times for 37 satellites of the MW, which ranged from $\approx 1 -11 \Gyr$ ago, similar to other simulation-focused studies \citep[for example][]{Wetzel15, Bakels21, Santistevan23}.
Deriving these infall and orbit history properties is generally difficult, given that we do not know precisely how the mass distribution of the MW or M31 has changed over time.


Stellar streams arise from the disruption of satellite galaxies or globular clusters.
Therefore, studying the orbits of streams or globular clusters gives insight into the possible future orbits of satellites that will eventually merge into their host galaxy \citep[for example][]{Ibata94, Majewski96, Bullock05, PriceWhelan16, PriceWhelan19, Panithanpaisal21, Bonaca21, Ishchenko23}.
Several studies used both the Dark Energy Survey \citep[DES;][]{DES} and the Southern Stellar Stream Spectroscopic Survey \citep[$S^5$;][]{Li19} to discover these small systems and measure their kinematics \citep[for example][]{Shipp18, Shipp19, Li21, Li22}.
Comparisons with cosmological simulations in \citet{Shipp23} suggest that undetected stellar streams may exist around the MW, which the upcoming Vera Rubin Observatory could potentially discover.

Cosmological simulations of MW-mass galaxies allow us to study theoretically the orbital evolution of satellites.
Many studies used DMO simulations to understand how subhalos respond to pericentric events \citep{Robles21}, how subhalo orbits respond to various MW environments \citep{Penarrubia02, Penarrubia05, Ogiya21} and pre-processing and group accretion \citep{Rocha12, Wetzel15, Li20_infall, Bakels21}.
However, many such studies did not account for the important effects of baryons \citep[for example][]{Brooks14, Bullock17, Sales22}.

Of utmost importance in deriving the satellite orbit histories is understanding the mass distribution of the MW and M31.
Studies such as \citet{Bovy13} and \citet{Bovy16} focused on fitting and deriving parameters for the disc, such as the scale height and length, while other studies estimated the total mass of the MW or M31 \citep[for example][]{Eadie17, Patel18, Eadie19, Patel23}.
One method of defining the total virial mass of a galaxy is by summing the mass within a given radius, such as $\Rthm$, the radius that encompasses $200\times$ the matter density of the Universe \citep{Bryan98}.
Using constraints from globular cluster kinematics, \citet{Vasiliev19} found that the virial mass of the MW is $\Mthm = 1.2 \times 10^{12} \Msun$, which is in line with the studies mentioned above and with the currently accepted virial mass in the literature of $\Mthm = 1 - 2 \times 10^{12} \Msun$ \citep[for example][]{BlandHawthorn16}.
Many studies use the orbits of satellite galaxies to constrain the mass of the MW or M31 \citep[for example][]{Evans00, vanderMarel08, Watkins10, Irrgang13}.
\citet{Patel23} suggested the mass of M31 to be more massive, $\Mthm = 2.85 - 3.02 \times 10^{12} \Msun$, from proper motions from HST and Gaia of satellite galaxies.

Many studies used numerical tools to backward integrate the orbits of satellites, stellar streams, or globular clusters, such as \galpy \ \citep{Bovy15}, \textsc{AGAMA} \citep{AGAMA}, and \textsc{Gala} \citep{gala}.
However, such orbit modeling often makes approximations by keeping the host mass profile fixed over time \citep[for example][]{Patel17, Fritz18, Fillingham19, Pace22}, sometimes varying the MW center of mass, including an LMC-mass satellite, or including dynamical friction \citep{Weinberg86, Lux10, Gomez15, Patel20, GaravitoCamargo19, GaravitoCamargo21, CorreaMagnus22, Lilleengen23}.

\citet{Lux10} compared various orbit history properties of subhalos in the dark matter-only Via Lactea I simulation \citep{Diemand07} to the orbits of MW satellites by using proper motion measurements from the literature and integrating their orbits in fixed potentials.
In another study, \citet{Arora22} compared 4 models with and without time dependence to investigate the effects of different mass models on stellar streams dynamics in simulations, and found that although most models conserve stream orbit stability, the only model that conserves stability over long periods of time is the time-evolving model.
\citet{DSouza22} used 2 MW-mass host halos from the ELVIS suite of dark-matter-only (DMO) simulations to test how well orbit modeling recovers the cosmological orbits of subhalos.
Although the majority of dynamical models applied to the MW and M31 assume static host potentials, the fiducial model in \citet{DSouza22} accounted for the true mass growth of each MW-mass host.
They compared results from host halos with and without LMC-mass satellites and showed that orbit modeling better recovers the more recent pericentres and apocentres when compared to the second or third-most recent.
They also tested models in which they did not account for any mass growth of the MW-mass host, or the presence of an LMC-like satellite, and found varying degrees of uncertainty associated with each simple model.
However, these simulations lacked baryonic physics, including the gravitational effects of a central galaxy, and various works noted the importance of modeling baryonic physics on these scales \citep[for example][]{Brooks14, ElBadry16, Bullock17, Sales22}.


In \citet{Santistevan23}, we studied the orbital dynamics and histories of satellite galaxies in the FIRE-2 cosmological zoom-in simulations of MW-mass galaxies.
We investigated trends between the present-day dynamical properties, such as velocity, total energy, and specific angular momentum, with the satellite infall times, present-day distance from the MW-mass host, and satellite stellar mass.
We also similarly checked for trends with properties at pericentre and found that the most recent pericentre was not the smallest, contrary to the expectation that satellite orbits only shrink over time.

In this paper, we further study the infall and orbital histories of the same satellites.
We model an axisymmetric mass profile for each simulated MW-mass host to within a few per cent at $z = 0$, and we backward integrate the orbits of satellites within each one.
We then compare these results against the `true' orbital histories of satellite galaxies in the simulations.
Our goal is to quantify rigorously the strengths and limitations of modeling satellite orbits in a static host halo potential, a commonly used technique.
Although we focus on satellite galaxies, our results are relevant for orbit models of stellar streams and globular clusters.

Key questions that we address are:
(1) How much has the mass profile of a MW-mass host evolved over the orbital histories of typical satellites?
(2) How well does orbit modeling in a static axisymmetric host potential recover key orbital properties in the history of a typical satellite?
(3) How far back in time can one reliably model the orbital history of satellites in a static axisymmetric host potential?


\section{Methods}   %
\label{sec:methods} %

\subsection{FIRE-2 simulations} %
\label{sec:sims}                %

\begin{table}
\centering
\begin{threeparttable}
\caption{
Properties at $z = 0$ of the 13 MW/M31-mass galaxies/halos in the FIRE-2 simulations that we analyse, ordered by decreasing stellar mass.
Simulations with `m12' names are isolated galaxies from the Latte suite, while the others are from the `ELVIS on FIRE' suite of Local Group-like pairs.
Columns: host name; $M_{\rm star,90}$ is the host's stellar mass within $R_{\rm star,90}$, the disc radius enclosing 90 per cent of the stellar mass within 20 kpc; $\Mthm$ is the halo total mass; $\Rthm$ is the halo radius; and $N_{\rm satellite}$ is the number of satellite galaxies at $z = 0$ with $\Mstar > 3 \times 10^4 \Msun$ that ever orbited within $\Rthm$, totalling 493 across the suite.
}
\begin{tabular}{|c|c|c|c|c|c|c|c|c|}
\hline
\hline
Name & $M_{\rm star,90}$ & $\Mthm$ & $\Rthm$ & $N_{\rm satellite}$ & Ref \\
& [$10^{10} \Msun$] & [$10^{12} \Msun$] & [kpc]   & & \\
\hline
m12m    & 10.0 & 1.6 & 371 & 47 & A \\
Romulus & 8.0  & 2.1 & 406 & 57 & B \\
m12b    & 7.3  & 1.4 & 358 & 32 & C \\
m12f    & 6.9  & 1.7 & 380 & 44 & D \\
Thelma  & 6.3  & 1.4 & 358 & 34 & C \\
Romeo   & 5.9  & 1.3 & 341 & 36 & C \\
m12i    & 5.5  & 1.2 & 336 & 27 & E \\
m12c    & 5.1  & 1.4 & 351 & 41 & C \\
m12w    & 4.8  & 1.1 & 319 & 39 & F \\
Remus   & 4.0  & 1.2 & 339 & 36 & B \\
Juliet  & 3.3  & 1.1 & 321 & 40 & C \\
Louise  & 2.3  & 1.2 & 333 & 34 & C \\
m12z    & 1.8  & 0.9 & 307 & 26 & C \\
\hline
Average & 5.5  & 1.4 & 348 & 38 & - \\
\hline
\hline
\end{tabular}
\label{tab:hosts}
\begin{tablenotes}
\item \textit{Note:} Simulation introduced in: A: \citet{Hopkins18}, B: \citet{GarrisonKimmel19b}, C: \citet{GarrisonKimmel19a}, D: \citet{GarrisonKimmel17}, E: \citet{Wetzel16}, F: \citet{Samuel20}.
\end{tablenotes}
\end{threeparttable}
\end{table}

We use the cosmological zoom-in baryonic simulations of MW-mass galaxies in both isolated and LG-like environments from the Feedback In Realistic Environments (FIRE) project\footnote{See the FIRE project web site: \url{http://fire.northwestern.edu}} \citep{Hopkins18}.
We ran these simulations using the hydrodynamic plus $N$-body code \textsc{Gizmo} \citep{Hopkins15}, with the mesh-free finite-mass (MFM) hydrodynamics method \citep{Hopkins15}, and the FIRE-2 physics model that includes several radiative heating and cooling processes such as Compton scattering, Bremsstrahlung emission, photoionization and recombination, photoelectric, metal-line, molecular, fine-structure, dust-collisional, and cosmic-ray heating across temperatures $10 - 10^{10} \K$ \citep{Hopkins18}.
The FIRE-2 physics model also includes the spatially uniform and redshift-dependent cosmic ultraviolet (UV) background from \cite{FaucherGiguere09}, for which HI reionization occurs at $z_{\rm reion} \approx 10$.
Stars form in gas that is self-gravitating, Jeans unstable, molecular \citep[following][]{Krumholz11}, and dense ($n_H > 1000$ cm$^{-3}$), and represent single stellar populations, assuming a \cite{Kroupa01} initial mass function.
Stars then evolve along stellar population models from \textsc{STARBURST99 v7.0} \citep{Leitherer99}, inheriting masses and elemental abundances from their progenitor gas cells.
Other stellar feedback processes we implement in the FIRE-2 simulations include core-collapse and white-dwarf (Type Ia) supernovae, stellar winds, and radiation pressure.

We generated cosmological zoom-in initial conditions at $z \approx 99$ within periodic cosmological boxes of comoving length $70.4 - 172 \Mpc$, which are large enough to avoid unrealistic periodic gravity effects on individual MW-mass hosts, using the code \textsc{MUSIC} \citep{Hahn11}.
We saved 600 snapshots for each simulation with time spacing of $\approx 25 \Myr$ down to $z = 0$, assuming a flat $\Lambda$CDM cosmology with the following cosmological parameters consistent with \citet{Planck18}: $h = 0.68 - 0.71$, $\sigma_{\rm 8} = 0.801 - 0.82$, $n_{\rm s} = 0.961 - 0.97$, $\Omega_{\Lambda} = 0.69 - 0.734$, $\Omega_{\rm m} = 0.266 - 0.31$, and $\Omega_{\rm b} = 0.0449 - 0.048$.

We analyse a similar set of galaxies as \citet{Santistevan23}, only we omit `m12r', because of its low stellar mass compared to the MW and because we are not able to fit its mass profile to sufficiently high precision (see Appendix~\ref{app:mass_profile}).
We also include `m12z', first introduced in \citet{GarrisonKimmel19a}.
Our sample is from both the Latte suite of isolated MW/M31-mass galaxies, introduced in \citet{Wetzel16}, and the `ELVIS on FIRE' suite of LG-like MW+M31 pairs, introduced in \citet{GarrisonKimmel19a}.
Table~\ref{tab:hosts} lists several of their properties at $z = 0$, such as stellar mass, $M_{\rm star,90}$, halo mass, $\Mthm$, and radius, $\Rthm$, and the number of satellite galaxies at $z = 0$ with $M_{\rm star} > 3 \times 10^4 \Msun$, $N_{\rm satellite}$.

The Latte suite of isolated MW/M31-mass galaxies includes halos at $z = 0$ with $\Mthm = 1 - 2 \times10^{12} \Msun$ with no other halos of similar mass within $5 \Rthm$.
We also chose m12w to have LMC-mass satellite analogs near $z \approx 0$, and m12z to have a smaller DM halo mass at $z = 0$ \citep{Samuel20}.
Star particles and gas cells are initialised with masses of $7100 \Msun$, however, because of stellar mass loss, the typical is $\approx 5000 \Msun$.
The mass of dark-matter (DM) particles is $3.5 \times 10^4 \Msun$ within the zoom-in region.
The gravitational softening lengths for star and DM particles are fixed at 4 and 40 pc (Plummer equivalent), respectively, comoving at $z > 9$ and physical thereafter.
The gas cells use adaptive force softening, consistent with their hydrodynamic kernel smoothing, down to 1 pc.

The selection criteria for each pair of halos in the `ELVIS on FIRE' suite of LG-like galaxies is based on their individual masses ($\Mthm = 1 - 3 \times 10^{12} \Msun$), combined masses ($M_{\rm tot} = 2 - 5 \times 10^{12} \Msun$), their relative separation ($600 - 1000 \kpc$) and radial velocities ($\upsilon_{\rm rad} < 0 \kmsi$) at $z = 0$.
The mass resolution in the `ELVIS on FIRE' suite is $\approx 2 \times$ better than the Latte suite, with initial masses of star particles and gas cells of $\approx 3500 - 4000 \Msun$.

Our 13 simulated galaxies display broadly consistent properties as similar MW/M31-mass galaxies and exhibit comparable observational properties to the MW or M31, such as: MW/M31-like morphologies \citep{Ma17, GarrisonKimmel18, ElBadry18, Sanderson20} that follow stellar-to-halo mass relations \citep{Hopkins18}, realistic stellar halos \citep{Bonaca17, Sanderson18}, and dynamics of metal-poor stars from early galaxy mergers \citep{Santistevan21}.
Each galaxy also hosts a satellite galaxy population with properties comparable to the satellites within the local Universe, such as: stellar masses and internal velocity dispersions \citep{Wetzel16, GarrisonKimmel19b}, radial and 3-D spatial distributions \citep{Samuel20, Samuel21}, star-formation histories and quiescent fractions \citep{GarrisonKimmel19b, Samuel22}.

\subsection{Halo/galaxy catalogs and merger trees} %
\label{sec:rockstar}                               %

To generate the (sub)halo catalogs at each of the 600 snapshots, we use the \textsc{ROCKSTAR} 6D halo finder \citep{Behroozi13a} using DM particles only, and we use \textsc{CONSISTENT-TREES} \citep{Behroozi13b} to generate merger trees.
None of the (sub)halos that we analyse have any low-resolution DM particle contamination, given the sufficiently large zoom-in volumes.

We briefly review how we implement star particle assignment in post-processing here, but we refer the reader to \citet{Samuel20} for details.
First, we select star particles within $d < 0.8 \, R_{\rm halo}$, out to a maximum distance of $30 \kpc$, with velocities within $\upsilon < 2 \, V_{\rm circ,max}$ of the (sub)halo's center-of-mass (COM) velocity.
We then keep the star particles within $d < 1.5 \, R_{\rm star,90}$ of the (then) current member stellar population's COM and (sub)halo center position, where $R_{\rm star,90}$ is the radius that encloses 90 per cent of the stellar mass.
Then, we select the star particles with velocities within $\upsilon < 2 \, \sigma_{\rm vel,star}$ of the COM velocity of the member star particles, where $\sigma_{\rm vel,star}$ is the velocity dispersion of the current member star particles.
Finally, we iterate on both the spatial and kinematic criteria until the (sub)halo's stellar mass converges to within 1 per cent.
This also guarantees that the COM of the galaxy and its (sub)halo are consistent with one another.

We use two publicly available analysis packages: \textsc{HaloAnalysis}\footnote{\url{https://bitbucket.org/awetzel/halo\_analysis}} \citep{HaloAnalysis} for assigning star particles to halos and for reading and analyzing halo catalogs/trees, and  \textsc{GizmoAnalysis}\footnote{\url{https://bitbucket.org/awetzel/gizmo\_analysis}} \citep{GizmoAnalysis} for reading and analyzing particles from Gizmo snapshots.

\subsection{Selection of satellites} %
\label{sec:selection}                %

We select satellites in the same manner as \cite{Santistevan23}.
To summarise, we include all satellites at $z = 0$ with $\Mstar > 3 \times 10^4 \Msun$ that ever orbited within their MW-mass halo's virial radius, $\Rthm$.
This stellar mass limit corresponds to roughly $\approx 6$ star particles, which reasonably resolves the total stellar mass \citep{Hopkins18}.
At our selection threshold of $\Mstar > 3 \times 10^4 \Msun$, the median peak halo mass is $M_{\rm halo,peak} \approx 9 \times 10^8 \Msun$ which corresponds to $\gtrsim 2 \times 10^4$ dark-matter particles.
Thus, we resolve satellite subhalos well, to prevent significant numerical disruption according to the criteria in \citet{vandenBosch18}; see \citet{Samuel20} for more discussion on our satellite resolution convergence.

We include `splashback' satellite galaxies that currently orbit outside of the MW-mass halo's $\Rthm$ but are gravitationally bound to it \citep[for example][]{Wetzel14}.
As Table~\ref{tab:hosts} shows, the number of surviving satellites at $z = 0$ per host, including the splashback population, is $26 - 57$, and our sample totals 493 satellites.

To avoid biasing our results to the hosts with more satellites, we oversample the satellites, so that each host contributes a near equal fraction of satellites to the total, to within 5 per cent; see \cite{Santistevan23} for details.

\subsection{Calculating orbit properties} %
\label{sec:pipeline}                      %

Many dynamical modeling studies implement a static gravitational potential for the host, consisting of a sum of potentials for each component of the galaxy, such as the stellar/gaseous disc, the bulge, the stellar halo, and the dark-matter halo \citep[see for example][]{Kallivayalil13, Gomez15, Patel17}.
To numerically integrate orbits through time, these studies then often use common numerical tools, such as \textsc{Galpy}, to solve the equations of motion at each timestep.

In our analysis, we backward integrate the orbits of satellite galaxies in mass profiles that we fit to each MW-mass host in the FIRE-2 simulations.
In short, we model the mass profiles of the hosts at $z = 0$ with a generalized form of the spherical Navarro-Frenk-White \citep[NFW,][]{NFW} density profile using dark matter and hot gas ($T > 10^5 \K$) particles within $r < 10 \kpc$, and \textit{all} particles at $r = 100-500 \kpc$.
We model the disc with two double-exponential disc profiles, one for the inner disc (bulge) and one for the outer disc, using star particles and cold gas ($T < 10^5 \K$).
The median fit across all 13 MW-mass hosts is within $\approx 5$ per cent of the enclosed mass profiles in the simulations at $r > 10 \kpc$ out to the virial radius.
Thus, \textit{we test orbit modeling under a best-case scenario, at least for a static axisymmetric potential, with near perfect knowledge of mass profile at present day}.
We note that we do not model the additional gravitational potential for any given satellite galaxy.
See Appendix~\ref{app:mass_profile} for more details on our fits to each MW-mass host.

Next, we select the cylindrical positions ($R, \phi, Z$) and velocities ($v_{\rm R}, v_{\rm \phi}, v_{\rm Z}$) of the satellites at the $z = 0$ snapshot and use this 6D phase-space information to initialise their orbits.
We then use the galactic dynamics python package \galpy\footnote{\url{http://github.com/jobovy/galpy}} \citep{Bovy15} to backward integrate the satellite orbits for $13.8 \Gyr$ within each host's static axisymmetric potential.
Because the MW-mass host is the only gravitational potential we account for, we do not include any movement of the MW-mass host throughout the satellite orbit integration.
This paper focuses on understanding the base uncertainties in static potential orbit modeling, thus, we do not account for dynamical friction in our model.
Including dynamical friction may improve the model orbits, however this is outside of the scope of this paper.

We explore numerous properties of satellite orbits, each of which provides insight into their orbit histories.
For example, pericentres occur when the satellite is at its closest approach to the MW-mass host, when the satellite feels the strongest tidal forces and is deepest in the host CGM.
Some studies use the first post-infall apocentre distance, also called the turn-around radius, as an alternate definition for the radius of the host galaxy \citep[for example][]{More15, Diemer17}.
The orbital eccentricity of satellites describes the orbit shape, which can change over time in the simulations, but is fixed in the fixed potential models.
The orbital energy is invariant in a time-independent potential, and in a spherically symmetric potential, total angular momentum is invariant.
Thus, comparing the evolution of these properties with the simulations informs us to what extent this holds.
We calculate the following properties for the satellites in our sample.

\textit{Pericentre distance, time, velocity, and number}:
We define the virial properties at $\Rthm$, the radius that encloses $200 \times$ the mean matter density of the Universe.
We calculate pericentres in the same manner as \citet{Santistevan23}.
First, we track the main progenitor of the satellites back in time through all 600 snapshots using the merger trees, and first confirm that the satellite is within the virial radius, $\Rthm$, of the MW-mass host halo at a given snapshot.
Then, we find local minima in its galactocentric distance within a $\pm 20$ snapshot window, which corresponds to $\approx 1 \Gyr$ in time.
Given the $\approx 25 \Myr$ time spacing between snapshots, we then fit a cubic spline to the distance, time, and velocity arrays in this snapshot window, and save the spline interpolated minimum distances, and the corresponding times and velocities at these pericentres.
Finally, we assign the total number of pericentres to a satellite based on the number of times the above criteria are met.
As mentioned in \citet{Santistevan23}, we checked our pericentre calculations using a snapshot window of $\pm 4, 8, 10$ snapshots and saw that our fiducial window of $\pm 20$ snapshots best eliminates `false' pericentres, that is, cases where the pipeline finds a pericentre in either numerical noise or short-lived perturbations in the orbits.
Additionally, whether we center the distances on the satellite or MW-mass host galaxies or DM (sub)halos does not affect our results.
    
\textit{Apocentre distance}:
We calculate apocentres similar to the way that we calculate pericentres.
We first confirm that the satellite has orbited within $\Rthm$ of the MW-mass host halo before a given snapshot, however, we do not require it to be within $\Rthm$ at the snapshot of interest so that we may catch apocentres in the `splashback' phase.
We then find local maxima in the satellite's galactocentric distance within a similar window of $\pm 20$ snapshots.
Finally, we similarly fit a cubic spline to the distance and save these values.
    
\textit{Infall time}:
To calculate infall time, we simply ensure that the satellite is within $\Rthm$ at a given snapshot, and save the corresponding time that this first happens.
In orbit modeling, we calculate infall time in two different ways depending on how we treat $\Rthm$.
The first method involves using the evolving $\Rthm$ from the simulations, and finding when the model orbit first crosses this distance, similar to how we calculate infall time for the satellites in the simulations.
However, in our second method, we keep the present-day $\Rthm$ as a fixed quantity over time, that is, $\Rthm(t_{\rm lb} = 0)$, where $\tlb$ is lookback time.
We find the instances in which the model orbit first crosses within this fixed distance.
In the case that the model orbit was always within $\Rthm(t_{\rm lb} = 0)$, we set the infall time equal to the beginning of the simulations, $13.8 \Gyr$ ago.
    
\textit{Eccentricity}:
We approximate orbital eccentricity as:
\begin{equation}
e \approx \frac{d_{\rm apo} - d_{\rm peri}}{d_{\rm apo} + d_{\rm peri}}
\end{equation}
where $d_{\rm apo}$ and $d_{\rm peri}$ are the apocentre and pericentre distances, respectively.
Defined this way, the eccentricity is a constant for a Keplerian orbit.
However, it will vary within our model here, thus we choose adjacent pairs of apocentres and pericentres in the actual integrated orbit from the model to calculate it.
We make no distinction in the ordering of pericentre/apocentre combinations, i.e. whether not to choose an apocentre only after pericentre, or pericentre only after an apocentre.
    
\textit{Orbital period}:
We approximate the orbital period by simply calculating the time between adjacent pericentres.
We checked how this compares to the timing between adjacent apocentres and found consistent results.
    
\textit{Specific orbital energy}:
We take the specific orbital energy of a satellite to be the sum of the kinetic and potential energy per mass at each snapshot.
The simulation snapshots store the gravitational potential at the location of each particle, which we use to compute at the location of a satellite.
Following \citet{Santistevan23}, we select all star, gas, and dark matter particles within $\pm 5 \kpc$ of the satellite's virial radius, and use the median potential of these particles, to minimize the satellite's self-potential.

Because we track satellites across time, we must properly normalise the potentials at each snapshot.
Our sample includes 3 LG-like pairs of MW/M31-mass hosts, thus, we cannot normalise the potentials at arbitrarily large distances.
Therefore, we choose to normalise potentials with:
\begin{multline}
U_{\rm sat}(r, \tlb) = U_{\rm sat, snap}(r, \tlb) - U_{\rm host, snap}(r=500\kpc, \tlb) \\ - \frac{G \times M(< 500 \kpc, \tlb = 0)}{500 \kpc} + \frac{G \times M(< 500 \kpc, \tlb)}{500 \kpc}
\label{eq:norm}
\end{multline}
where $U_{\rm sat, snap}(r, \tlb)$ is potential of a satellite at a given snapshot, $U_{\rm host, snap}(r = 500\kpc, \tlb)$ is the potential for particles within a spherical shell at $r = 500 \pm 5 \kpc$ around the MW-mass host, and the last two terms are the analytic gravitational potentials, $G \times M(< r) / r$, for the mass enclosed within $500 \kpc$ at present-day and any given lookback time, respectively.
We choose $500 \kpc$ for several reasons: (i) the bound mass for a given MW-mass host does not change by more than a few per cent beyond this, (ii) satellites typically orbit as far as the `splashback' radius, which we approximate as $\approx 1.5 \, \Rthm \approx 500 \kpc$ from a spherical collapse model \citep[for example][]{Fillmore84, Bertschinger85}, and (iii) we must choose distances smaller than the separation between the two MW-mass hosts ($\gtrsim 840 \kpc$).

For the analytic potentials, we get
\begin{equation}
\Phi = \int_{500\kpc}^{\infty} \frac{G\times M(<r)}{r^2} dr
\end{equation}
and because the enclosed mass does not change significantly at $500 \kpc$, this integral results in $\Phi \approx -G \times M(< 500 \kpc) / 500 \kpc$.
The last three terms in Equation~\ref{eq:norm} ensure that the potential is properly normalised across different snapshots.

When we examine differences in total energy, we divide by the virial potential of the host halo, $U_{\rm 200m,0} \equiv G \times M(r < \Rthm, \tlb = 0) / \Rthm$, because each host has different $\Mthm$ and $\Rthm$, so this ensures that we compare satellite evolution in a similar manner.
However, the snapshots for `m12z', `Romulus', and `Remus' do not have stored potential values, thus, we exclude them when we compare orbital energies.

\textit{Specific angular momentum}:
We calculate a satellite's specific angular momentum at each snapshot with $\ell = r \times v$, where $r$ is the total distance between the satellite and the center of the MW-mass host, and $v$ is the total velocity of the satellite with respect to the center of the MW-mass host.

\textit{Tidal acceleration}:
Finally, we calculate the tidal acceleration a satellite feels by taking the derivative of $a = G \times M(< r) / r^2$ with respect to $r$, where $M(< r)$ is the total enclosed mass of the host within a distance $r$.
We calculate this at every snapshot and save only the maximum $\dadr$ that a satellite experienced after first infall, although this almost always coincides with when the satellite is at its closest approach to the MW-mass host.

\subsection{Disc orientation} %
\label{sec:orientation}       %

We use the 6D phase-space coordinates of satellites at $z = 0$ relative to the MW-mass disc, but the disc precesses over time.
In \citet{Santistevan21}, we showed that after the disc stabilizes $\approx 5 - 11.5 \Gyr$ ago (when the angular momentum vector of the disc stopped rapidly fluctuating in its orientation), the disc continued to precess between $5 - 130^{\circ}$ until $z = 0$.
Satellites reach their closest approach to the MW-mass disc when they are at pericentre, so pericentre properties are likely most sensitive to the host disc configuration.
Thus, we explore different disc orientations and models to investigate how they affect the resultant satellite orbits in Appendix~\ref{app:point_mass}.
To summarise, in one model we rotate the disc by $90^{\circ}$ while keeping the same coordinates for satellites at $z = 0$.
In another model, we use a point mass model for the disc.
In both models, the median differences between all orbit properties we explore between our fiducial model and these different configurations is small, less than 1 per cent.
Therefore the details of the geometric configuration of the galactic potential do not matter much to satellite galaxy orbits.
For more discussion about this see Appendix~\ref{app:point_mass} and Table~\ref{tab:disc_model}.

\section{Results}   %
\label{sec:results} %

Many studies integrate satellite orbits using a model of a static axisymmetric MW/M31 potential at $z = 0$, which is unphysical given that the MW evolves over time.
Therefore, to provide context for our results on satellite orbits, we first quantify the mass evolution of MW-mass hosts in our simulations, over both long and short timescales, including how this depends on distance.
We then explore the extent to which satellites conserve energy and angular momentum.
Finally, we explicitly compare results between the simulations and the idealized axisymmetric model.
Because many of these distributions are non-Gaussian, throughout we present the median trends across the sample of host galaxies or satellites, as well as the half-width of the 68th or 95th percentile range, which for brevity we refer to as the $1 \sigma$ and $2 \sigma$ scatter, respectively.

\subsection{Growth of the Milky Way-mass host}
\label{sec:mp_evo}

\subsubsection{Halo virial properties} %

\begin{figure*}
\centering
\begin{tabular}{c}
\includegraphics[width = 0.95 \linewidth]{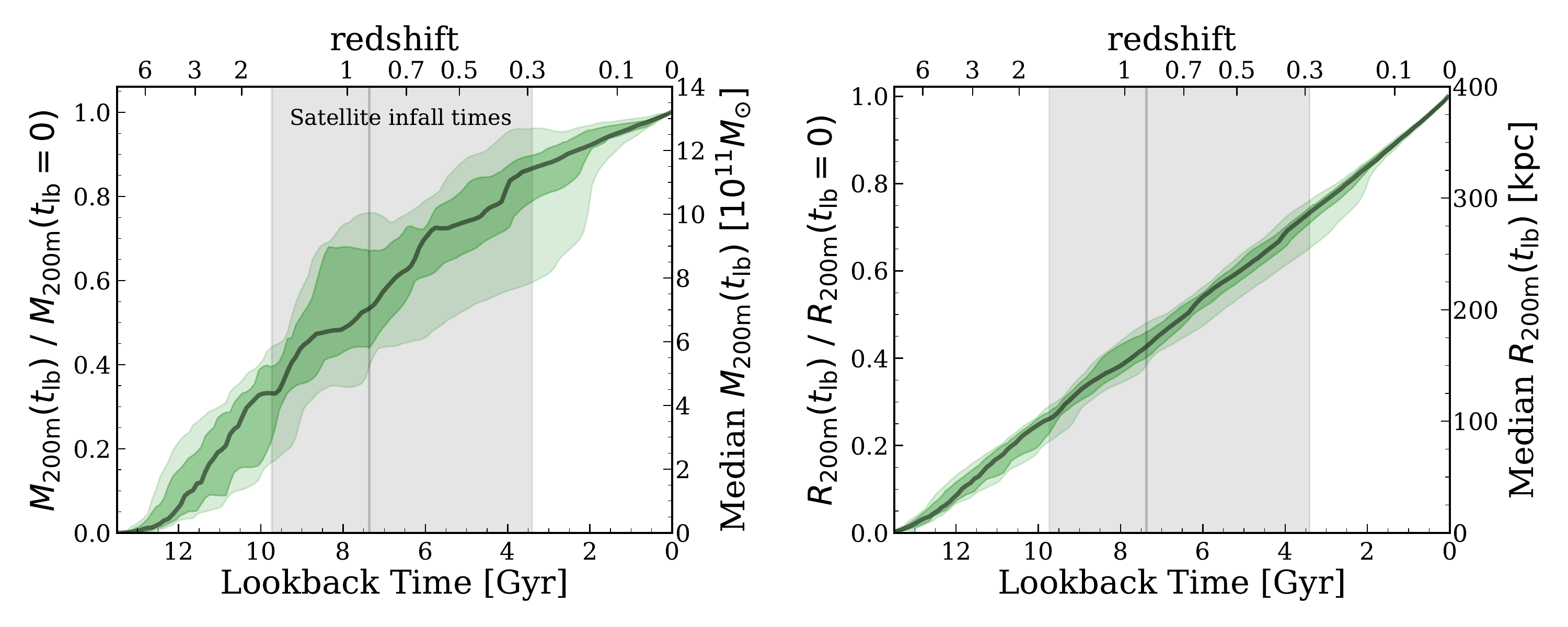}
\end{tabular}
\vspace{-2 mm}
\caption{
\textbf{Left}: Total (baryonic plus dark matter) mass of a MW/M31-mass halo, $\Mthm$, enclosed within $\Rthm$ as a function of lookback time, $\tlb$.
The left axis shows $\Mthm$ at $\tlb$ relative to $\Mthm$ today, while the right axis converts this to the median $\Mthm$ across our sample.
The black line shows the median, and the dark and light shaded regions show the 68th and 95th percentiles across our 13 hosts.
The vertical grey line and shaded region show the median and 68th percentile range of the lookback times of infall for surviving satellites.
The typical host halo had 54 per cent of its final $\Mthm$ when a typical satellite fell in $\approx 7.4 \Gyr$ ago, so $\Mthm$ nearly has doubled over that time.
\textbf{Right}: Same, for the growth of the MW/M31-mass halo radius, $\Rthm$.
$\Rthm$ shows nearly linear growth with time over the last $\approx 12 \Gyr$.
The typical host halo had 43 per cent of its final $\Rthm$ when a typical satellite fell in $\approx 7.4 \Gyr$ ago, again highlighting significant change over the orbital history of a typical satellite.
}
\label{fig:hosts}
\end{figure*}

We define virial properties at $\Rthm$, the radius that encloses $200 \times$ the mean matter density of the universe.
Figure~\ref{fig:hosts} shows the evolution of both the total (baryonic plus dark matter) virial mass of the hosts, $\Mthm$, and their virial radii, $\Rthm$.
We show the median trend along with the 68th and 95th percentiles.
The left-hand axes show the virial properties normalised to the present day ($\tlb = 0$), while the right-hand axes show these in physical units, scaling to the median.
The grey line and shaded region also show the median and 68th percentile of the infall times, $\MWinfall$, of the satellites in our sample.

The median MW-mass host $\Mthm$ grew more quickly at earlier times, slowing around $4 \Gyr$ ago.
As \citet{Santistevan20} showed, the fractional stellar mass growth of these MW-mass hosts is broadly consistent with studies based on abundance matching and dark-matter-only simulations \citep[for example][]{Behroozi13b, Hill17}.
Most satellites fell in $3.4 - 9.7 \Gyr$ ago, when the median MW-mass host had $\approx 33 - 86$ per cent of its mass at $z = 0$.

Figure~\ref{fig:hosts} (right) shows the growth of $\Rthm$ is nearly linear in time, with relatively small fractional scatter.
When the typical surviving satellites fell in, the median MW-mass host had $26 - 73$ per cent of its $\Rthm(z = 0)$.
Thus, the MW-mass hosts grew considerably in mass and radius, which affects the orbits of satellites.



\subsubsection{Mass within fixed physical radii} %
\label{sec:mp_long}                     %

\begin{figure}
\centering
\begin{tabular}{c}
\includegraphics[width = 0.95 \linewidth]{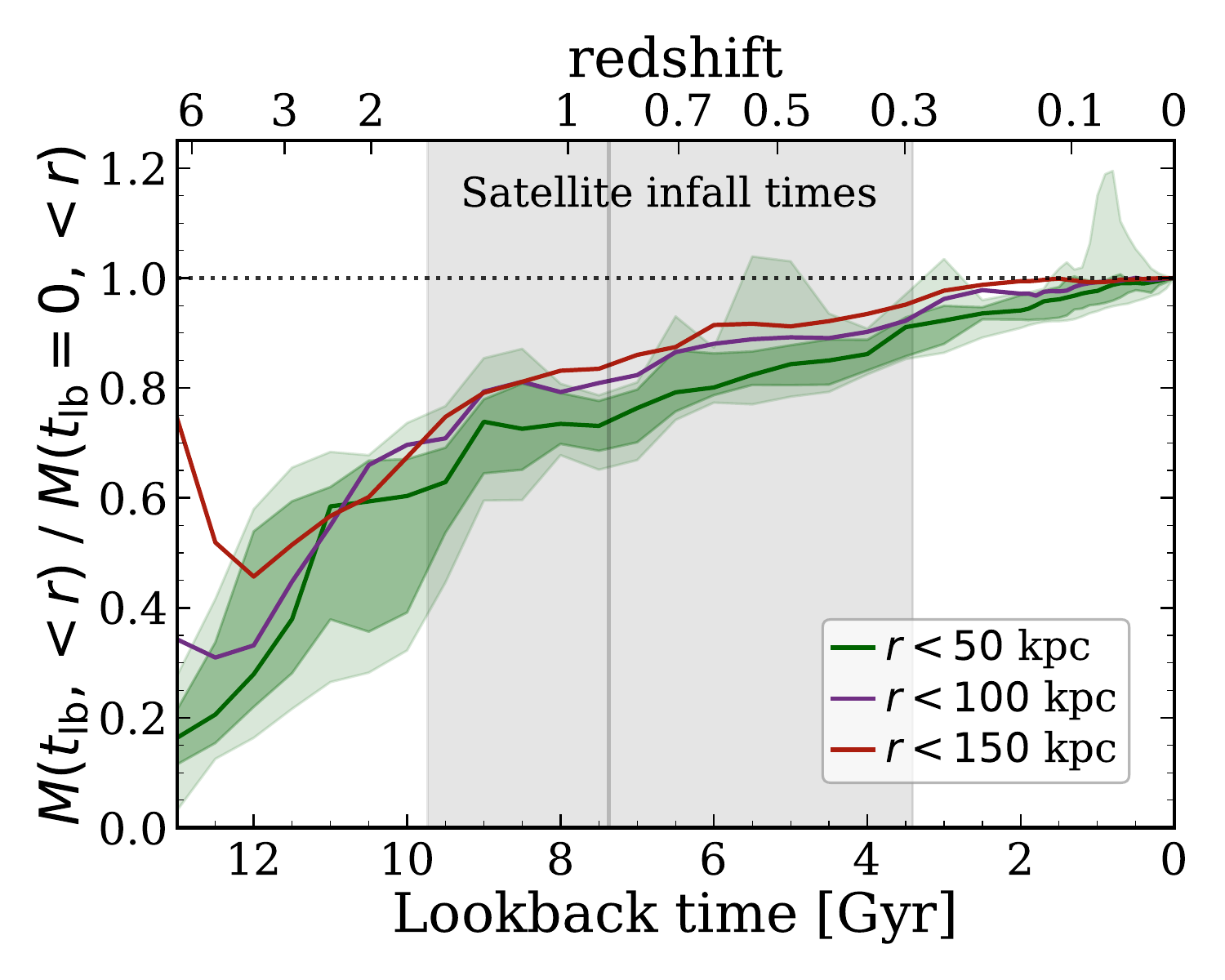}
\end{tabular}
\vspace{-2 mm}
\caption{
Total mass of the MW/M31-mass host within a fixed physical distance, $r$, normalised to the value today, versus lookback time, $t_{\rm lb}$.
We show the median mass ratio within $r = 50$, 100, and 150 kpc, and the 68th and 95th percentiles for $r < 50 \kpc$.
This `physical' mass growth is less significant than that of $\Mthm$ (based on an evolving $\Rthm$) in Figure~\ref{fig:hosts}.
However, \textit{because of baryonic gas cooling, the fractional mass growth of the host is larger at smaller distance.}
The mass growth reverses at $t_{\rm lb} \gtrsim 11 \Gyr$, when a given distance experienced its initial collapse from the Hubble expansion.
For context, the median pericentre distance experienced by surviving satellites is $\approx 50 \kpc$.
A typical satellite fell in $3.4 - 9.7 \Gyr$ ago (grey shaded region), when the typical enclosed mass was $\approx 61 - 93$ per cent of its value today.
}
\label{fig:mp_vs_lb}
\end{figure}

Figure~\ref{fig:hosts} showed the enclosed mass within $\Rthm(t)$.
However, for satellites that already fell into the MW-mass halo, the additional growth $\gtrsim \Rthm$, may not matter much, given that the orbits depend primarily on just the mass within an orbit.
Therefore, Figure~\ref{fig:mp_vs_lb} shows the ratio of the enclosed mass within a given fixed physical radius, $r$, at a given lookback time, $t_{\rm lb}$, relative to today, $M(< r, t_{\rm lb}) / M(< r, t_{\rm lb} = 0)$.
We show the median ratios of enclosed mass within $r < 50$, 100, and 150 kpc, along with the 68th and 95th percentiles for $r < 50 \kpc$.
Because the satellite orbits are sensitive to the enclosed mass, we choose to measure the enclosed mass within these distances because typical pericentre distances for the satellites in our sample are $\sim50\kpc$ and typical apocentres are $\sim200\kpc$.

Because we show enclosed mass within fixed physical radii, the increase with lookback time $\gtrsim 12 \Gyr$ ago represents the Hubble expansion, prior to the collapse within that radius.
The enclosed mass was $\approx 62 - 91$ per cent of its present value during the typical infall times of surviving satellites; larger than for $\Mthm$ in Figure~\ref{fig:hosts} during the same time range.
The spikes in the 95th percentile range likely arise from massive satellites.
The enclosed mass fractions within $r < 100 \kpc$ and $r < 150 \kpc$ were larger at nearly all lookback times, meaning that mass has grown fractionally less at larger radii.
In other words, the significant growth of the central galaxy, largely via gas cooling/accretion, leads to more significant mass grows at smaller radii.

Figure~\ref{fig:mp_vs_r} (top) shows the evolution of the enclosed mass profile across time from $r$ = 5 - 500 kpc, where $r$ is the satellite distance from the MW-mass host.
We first calculate the median profile over all 13 hosts at each distance and snapshot with $\approx 1 \Gyr$ time spacing; see the colorbar for the lookback time of a given mass ratio.
Then we normalise the curves for each snapshot to the median profile of the hosts at present day.

In general, at each $r$ the enclosed mass ratio increased over time, and likewise, at each time, the enclosed mass ratio increases with distance.
Similar to Figure~\ref{fig:mp_vs_lb}, the mass growth over time at large $r$ was not as significant compared to small $r$.
For example, typical recent pericentre distances for satellites in the simulations are $\approx 50 - 60 \kpc$, and compared to present day, the enclosed mass was only 74 per cent during typical satellite infall times.
The median present-day distance of the satellite galaxies in the simulations is around $175 \kpc$, and at $\tlb = 7.4 \Gyr$ ago, the enclosed mass was $\approx 83$ per cent of its mass at $z = 0$.
However, near the virial radius and beyond, the enclosed mass was already 97 per cent during typical satellite infall times.

\subsubsection{Short-term evolution of host mass} %
\label{sec:mp_short}                              %

\begin{figure}
\centering
\begin{tabular}{c}
\includegraphics[width = 0.95 \linewidth]{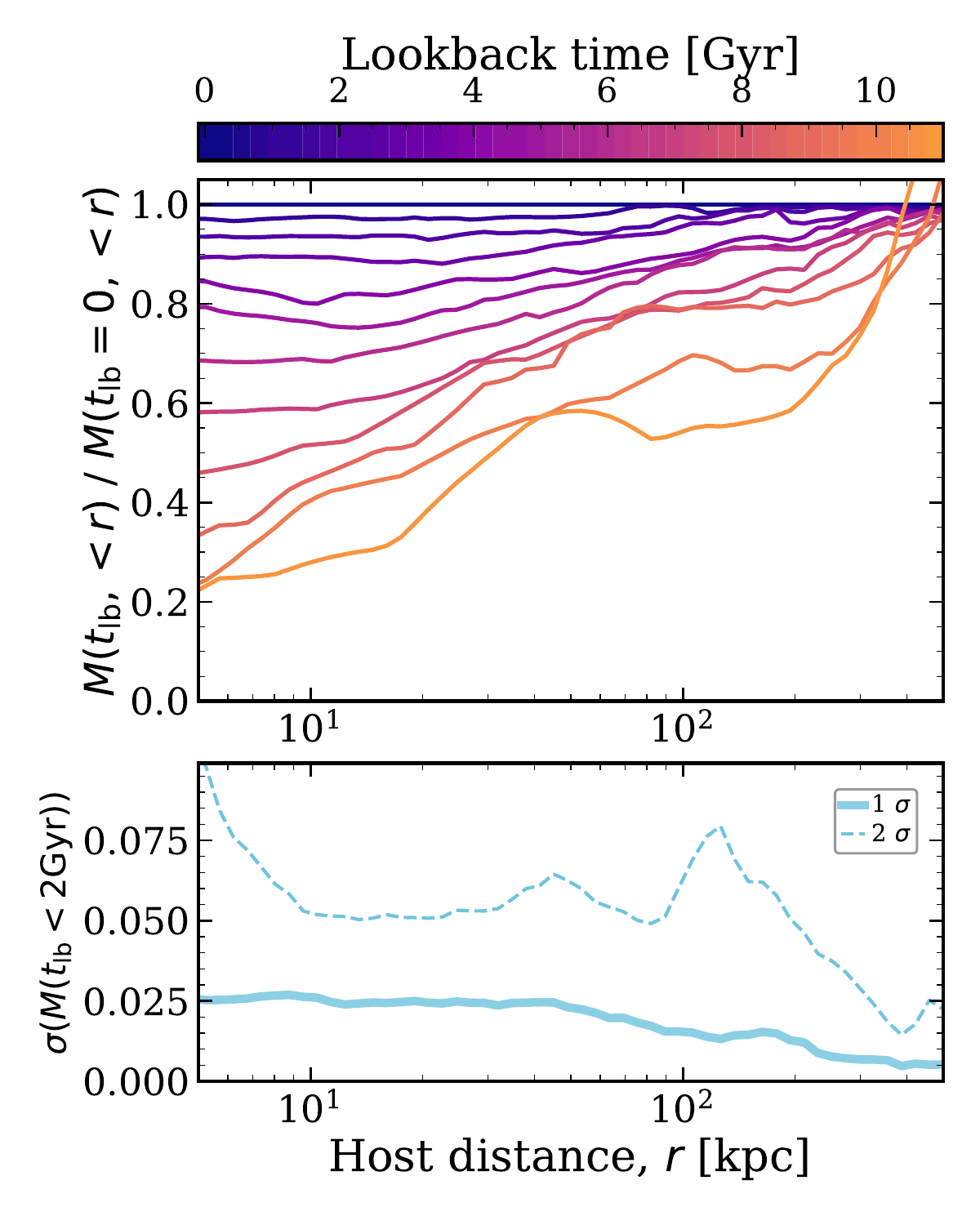}
\end{tabular}
\vspace{-2 mm}
\caption{
\textbf{Top: Average long-term evolution of the mass profile (Section~\ref{sec:mp_long}).}
Similar to Figure~\ref{fig:mp_vs_lb}, the total mass within a given physical distance, $r$, relative to the value today, but now as a function of $r$.
We show the median across our 13 MW/M31-mass hosts, at various lookback times, $t_{\rm lb}$, back to $11 \Gyr$, which encompasses the infall times of $> 95$ per cent of surviving satellites.
The enclosed mass increases over time at essentially all $r$, with the inner halo near the galaxy experiencing the most fractional mass growth.
Thus, the approximation of a static halo mass/potential is least accurate for satellites with the smallest pericentres.
For context, the median recent pericentre of surviving satellites is $\approx 50 \kpc$.
\textbf{Bottom: Typical short-term fluctuations in the host halo mass profile (Section~\ref{sec:mp_short}).}
The $1 \sigma$ and $2 \sigma$ standard deviation of the fractional change/fluctuation of the total mass of the host over the last 2 Gyr, which is the typical of the orbital timescale of satellites at $r \approx 30 \kpc$.
These fluctuations in enclosed mass are weaker at larger distances.
On average, the host mass growth/fluctuations over the last $2 \Gyr$ are $\lesssim 5$ per cent, though such fluctuations would be even higher for halos with a massive (LMC- or M33-mass) satellite.
}
\label{fig:mp_vs_r}
\end{figure}

The enclosed mass at any given time is subject to the perturbations of satellite galaxies that are orbiting around and merging within the main host.
To examine the variability in the enclosed masses, we calculate a time-averaged enclosed mass profile over the last 2 Gyr, and we normalise the enclosed mass at every snapshot between $\tlb = 0 - 2 \Gyr$ to this time-averaged one.
Figure~\ref{fig:mp_vs_r} (bottom) shows the $1 \sigma$ and $2 \sigma$ scatter of these ratios at each distance in the solid and dashed blue lines, respectively, in the bottom panel of Figure~\ref{fig:mp_vs_r}.


Similar to the top panel, the $1 \sigma$ scatter shows larger variability at small $r$, suggesting that the fractional growth in the inner regions of the MW-mass host is larger than at large $r$.
However, the $2 \sigma$ scatter does monotonically decrease with distance, but it is constant between $r = 10 - 100 \kpc$.

Halos and their galaxies grow hierarchically over time, and each figure in this section explicitly quantifies this idea in the evolving virial region (Figure~\ref{fig:hosts}), within fixed distance apertures (Figure~\ref{fig:mp_vs_lb}), and at fixed time (Figure~\ref{fig:mp_vs_r}).
Modeling the enclosed mass/potential of a host at $z = 0$ and holding that potential fixed across many Gyr does not account for the real, substantial growth of the host halo environment in which satellites orbit.

\subsection{Orbital energy and angular momentum} %
\label{sec:e_and_ell}                            %

\begin{figure*}
\centering
\begin{tabular}{c}
\includegraphics[width = 0.95 \linewidth]{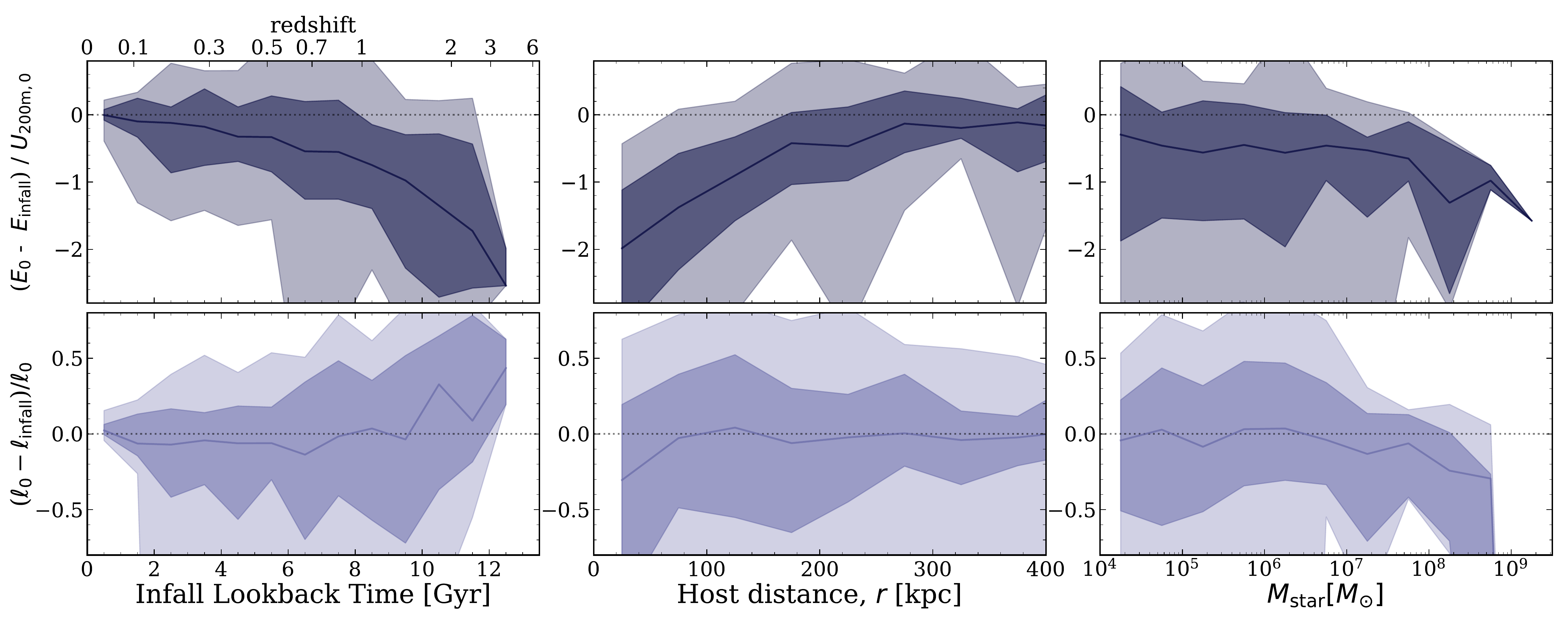}
\end{tabular}
\vspace{-2 mm}
\caption{
The difference in a satellite's specific orbital energy and specific orbital angular momentum between today and at first infall into the MW-mass halo, versus lookback time of infall (left), current distance from host, $r$ (middle), and satellite $\Mstar$ (right).
\textit{Orbit modeling nearly always assumes that these quantities are conserved.}
Solid lines show the median for all satellites within the 13 MW-mass hosts, and the dark and light shaded regions show the 68th and 95th percentiles.
\textbf{Top row}: Change in specific orbital total energy, $E$, relative to the virial energy of the MW-mass host halo today, $U_{\rm 200m, z = 0} = G \Mthm / \Rthm$.
The median decreases from $0$ to $-2.5$ with increasing infall lookback time, meaning that satellites have lost energy and become more bound since infall, because the host has grown (Figures~\ref{fig:hosts}-\ref{fig:mp_vs_r}).
Because $r$ inversely correlates with infall time, the median change in energy increases from $-2$ to $-0.2$ with $r$ (middle) and is constant with satellite mass at $\Mstar \lesssim 10^{7.5} \Msun$, but it decreases at higher mass because of dynamical friction.
Although $E$ correlates with these properties, the widths of the 68th percentiles span $\approx 1.2 - 1.5$, highlighting the significant variation (and thus uncertainty) for any given satellite.
\textbf{Bottom row}: Change in specific orbtal angular momentum, $\ell$, relative to the value today.
The median fractional difference is generally zero for satellites that fell in $\lesssim 9.5 \Gyr$ ago, but earlier infalling satellites have increased by up to $\sim 45$ per cent.
We find little to no change in the median $\ell$ with $r$ or $\Mstar$, except for satellites with $\Mstar \gtrsim 10^7 \Msun$, which experienced stronger dynamical friction.
The mean widths of the 68th percentiles in the fractional change in $\ell$ span $70$ per cent versus infall time and $r$, and 100 per cent versus $\Mstar$.
\textit{These uncertainties in both $E$ and $\ell$ imply that we cannot infer the initial energy or angular momentum of a given satellite's orbit at its time of infall to better than $\gtrsim 2$.}
}
\label{fig:e_ell_prop}
\end{figure*}

In a time-independent potential, the specific energy of a satellite's orbit is conserved.
Likewise, in a spherically symmetric potential, the specific angular momentum, $\ell = r \times v$, is conserved, while the component of $\ell$ along the minor axis is conserved in an axisymmetric potential.
In \citet{Santistevan23}, we showed that satellites that fell in $\lesssim 10 \Gyr$ ago conserve their median $\ell$ across the population, but importantly with large scatter of $\approx 40$ per cent.
However, we did not examine trends as a function of lookback time across an orbit.
We next examine how well orbital energy and angular momentum of a satellite's orbit are conserved.
We stress that we show trends across the full population of satellites, including the full range of values of satellites with a particular infall time, distance, and $\Mstar$, which gives a sense of conservation on a satellite-by-satellite basis.

Figure~\ref{fig:e_ell_prop} (top row) shows the difference in the total orbital energy between present-day and infall into the MW-mass halo.
To scale to the characteristic energy of a given halo, we divide these differences by $U_{\rm 200m,0} = G \Mthm / \Rthm$, the virial gravitational potential of the host halo today.
We show these fractional energy differences as a function of the lookback time of satellite infall into MW-mass host, $\MWinfall$, distance from the MW-mass host, $r$, and satellite stellar mass, $\Mstar$.

Generally, the median fractional difference in specific energy decreases with increasing $\MWinfall$ from 0 to $-2.5$ (top left).
Thus, satellites that fell into their MW-mass halo earlier lost more energy, because the MW-mass host grew in mass by $\gtrsim 20$ per cent (Figure~\ref{fig:mp_vs_lb}).
Next, the median fractional difference in total energy increases with $r$ (top middle), from $-2$ to roughly $-0.2$ for satellites near $\Rthm$.
Satellites that currently orbit at smaller distances have lost more energy since infall, compared to satellites that orbit at larger distances, which only recently fell in.
The fractional difference in total energy only weakly depends on $\Mstar$ (top right).
Satellites below $\Mstar \lesssim 10^{7.5} \Msun$ have a median fractional energy difference of roughly $-0.5$, and the fractional energy difference for more massive satellites decreases to as low as $-1.6$, but we caution that there is only one satellite with $\Mstar > 10^9 \Msun$.

Next, we investigate the specific angular momentum of satellite orbits, $\ell$.
My studies implement a spherically symmetric component to the potential, such as an NFW profile for the DM halo \citep[for example][]{Kallivayalil13, Patel17, Besla19}, in which the total angular momentum is constant.
We show in Appendix~\ref{app:point_mass} that satellite orbits are insensitive to the direction/details of the axisymmetric component of the potential, the disc, so we show results in $\ell$ and not the component of $\ell$ along the minor axis of the potential.

In \citet{Santistevan23}, we discussed trends in the angular momentum difference today versus infall, normalised by the angular momentum at satellite infall; $(\ell_0 - \ell_{\rm infall}) / \ell_{\rm infall}$.
Figure~\ref{fig:e_ell_prop} (bottom row) shows the same difference, only now normalised by the angular momentum at present-day, that is, $(\ell_0 - \ell_{\rm infall}) / \ell_0$.
Similar to \citet{Santistevan23}, we find weak dependence in the median fractional change in $\ell$ across the population since infall.
The median is as large as $\approx 45$ per cent compared to the values at infall.
These satellites make up roughly 20 per cent of the total sample.
The fractional difference in $\ell$ shows virtually no dependence with $r$ or $\Mstar$.
The average $1 \sigma$ scatter is $\approx 40$ and 50 per cent versus $r$ and $\Mstar$.

Our results suggest that satellite \textit{populations} do not show overall conservation of energy or angular momentum.
If we focus on the 68th and 95th percentile ranges in each panel, we see cases in which the energies or specific angular momenta for some satellites at present-day are similar to their values at infall, for a large range of infall time, $r$, and $\Mstar$.
However, this is not true for \textit{all} satellites, and the uncertainties in $E$ and $\ell$ suggest that \textit{one cannot determine the orbital energy or angular momentum of a given satellite at infall to within a factor of $\gtrsim 2$ from present-day measurements}.
Rather, satellites commonly lose some orbital energy since infall, likely because of the growth of the MW-mass host potential over time, the effects of dynamical friction on high-mass satellites, and also satellite-satellite interactions that may torque the satellite orbits.

We also investigate the extent to which the kinetic and potential energy components are conserved with respect to infall time, $r$, and $\Mstar$.
Versus infall time, the fractional change in the kinetic energy of satellites that fell in recently is positive and the fractional change in the potential energy was negative, suggesting these satellites are likely on their first infall and nearing their first pericentre.
Satellites with larger infall times often had negative fractional changes in kinetic and potential energies, because of both dynamical friction slowing the satellites and the growing host potential over time.
Within $r < 100 \kpc$, the circular velocity profile rises significantly, and satellites that orbit today have much larger kinetic energies compared to infall.
We note no strong trends with $\Mstar$.

\begin{figure}
\centering
\begin{tabular}{c}
\includegraphics[width = 0.95 \linewidth]{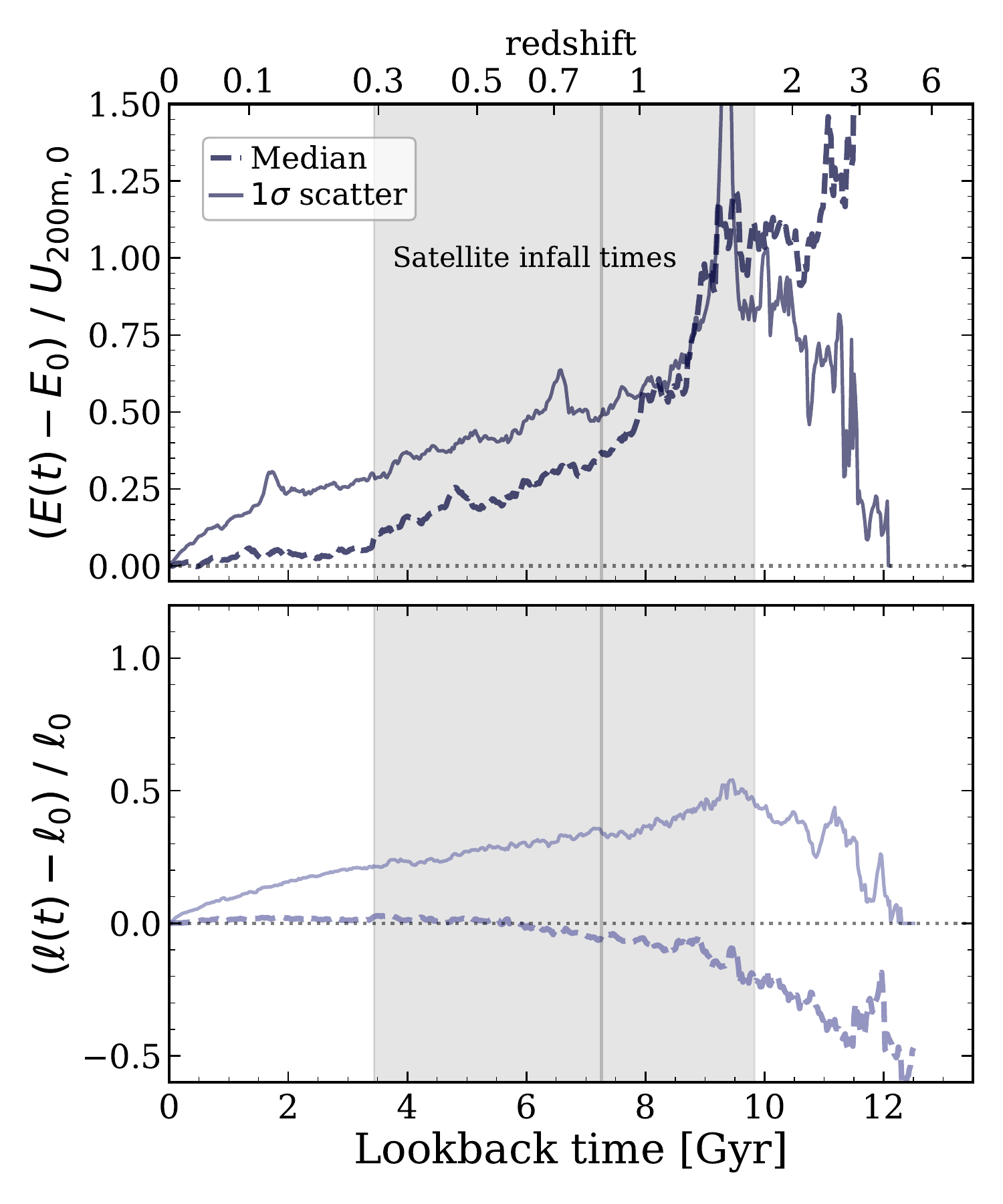}
\end{tabular}
\vspace{-2 mm}
\caption{
Fractional change in specific energy, $E$, and specific angular momentum, $\ell$, of satellite orbits versus lookback time.
Solid line shows the median across all satellites and dashed line shows the $1 \sigma$ scatter.
We only include galaxies that are satellites at a given lookback time, including splashback satellites.
The vertical line and shaded region represent the median satellite infall time and 68th percentile range of values, respectively.
\textbf{Top}: Specific energy.
The median fractional change increases with lookback time, that is, they were less bound at infall than they are today, because the host has grown (Figures~\ref{fig:hosts}-\ref{fig:mp_vs_r}).
While the median, which represents the systematic bias across the population, reaches 25 per cent at $4.8 \Gyr$ ago, the $1 \sigma$ scatter, which represents the uncertainty for a given satellite, reaches 25 per cent already at $1.6 \Gyr$ ago.
\textbf{Bottom row}: Specific angular momentum.
The median is about zero over the last $\approx 6 \Gyr$, but at earlier times the angular momentum decreases, down to $-60$ per cent, meaning that the specific angular momentum systematically has increased since infall.
Again, the $1 \sigma$ scatter reaches 25 per cent already $4.7 \Gyr$ ago.
Thus, while inferences of orbital energy and angular momentum, based on their conservation, are relatively \textit{unbiased} (for the overall population) over the last $\approx 5 - 8 \Gyr$, their \textit{uncertainties} for a given satellite are large already $\approx 1 \Gyr$ ago.
}
\label{fig:conservation}
\end{figure}

Having quantified the change in orbital energy and angular momentum from first infall to today, Figure~\ref{fig:conservation} quantifies the extent to which a satellite conserves $E$ and $\ell$ as a function of lookback time.
We show the median trend across the population of satellites in the dashed lines.
For simplicity, we present the $1 \sigma$ scatter across the entire population, at a given snapshot in the solid lines.
We include only galaxies that were still satellites at a given lookback time, including splashback satellites, so the size of the sample monotonically decreases with lookback time.

Figure~\ref{fig:conservation} (top) shows the difference in the satellite orbital energy between a given lookback time and today, $E(t) - E_{0}$, normalised by the MW-mass halo potential today, $U_{\rm 200m, 0}$.
Over the last $\approx 3.5 \Gyr$, the median total energy is relatively unchanged, but at earlier times, this fractional difference increases with increasing lookback time from 0 to as large as $\approx 2$ at $11.75 \Gyr$ ago.
The fractional change in $E$ reaches 25 per cent at $\approx 4.8 \Gyr$ ago, 50 per cent at $7.9 \Gyr$ ago, and 100 per cent at $9.2 \Gyr$ ago.

Although the median fractional energy change over the last $3.5 \Gyr$ is small, this is only a statement about the population, and it does not imply energy conservation for a typical satellite, given the large scatter.
The  $1 \sigma$ scatter reaches 25 per cent already at $\approx 1.6 \Gyr$ ago, 50 per cent at $6.1 \Gyr$ ago, and 100 per cent $9.1 \Gyr$ ago.

Figure~\ref{fig:conservation} (bottom) shows the fractional change in the specific angular momentum of satellites, $\ell$.
Similar to energy, we compute the difference of $\ell$ at each lookback time with the value today, but now we normalise this difference by $\ell$ today.
The median $\ell$ is constant for longer, over the last $\approx 6 \Gyr$, before which it decreases. 
This implies that early-infalling satellites gained angular momentum on average, as we showed in \citet{Santistevan23}.
The median fractional difference reaches 25 per cent at a much later time of $\approx 11.9 \Gyr$ ago, and 50 per cent at $11.4 \Gyr$ ago.
Compared to the fractional change in $E$, the $1 \sigma$ scatter reaches a given fraction later, 25 per cent at $\approx 4.2 \Gyr$ ago, 50 per cent at $9.3 \Gyr$ ago.
We stress again that, although the population median is conserved longer, this does not imply that a given satellite's $\ell$ is conserved for this long.
Rather, the $1 \sigma$ scatter represents the typical uncertainty for a given satellite.

Figures~\ref{fig:e_ell_prop} and \ref{fig:conservation} show that \textit{neither $E$ nor $\ell$ are conserved across time}, which agrees with Figures~\ref{fig:hosts}-\ref{fig:mp_vs_r}, and results from the growth and general time dependence of the host halo potential.
The $1 \sigma$ scatter represents the typical uncertainty for a given satellite, which is as large as $\approx 50$ per cent for $\ell$ around $9.3 \Gyr$ ago, and as large as a factor of 2 for $E$ at $9.1 \Gyr$ ago.

\subsection{Orbit modeling in static axisymmetric host potential} %
\label{sec:orbits}                                                %

\begin{figure*}
\centering
\begin{tabular}{c}
\includegraphics[width = 0.95 \linewidth]{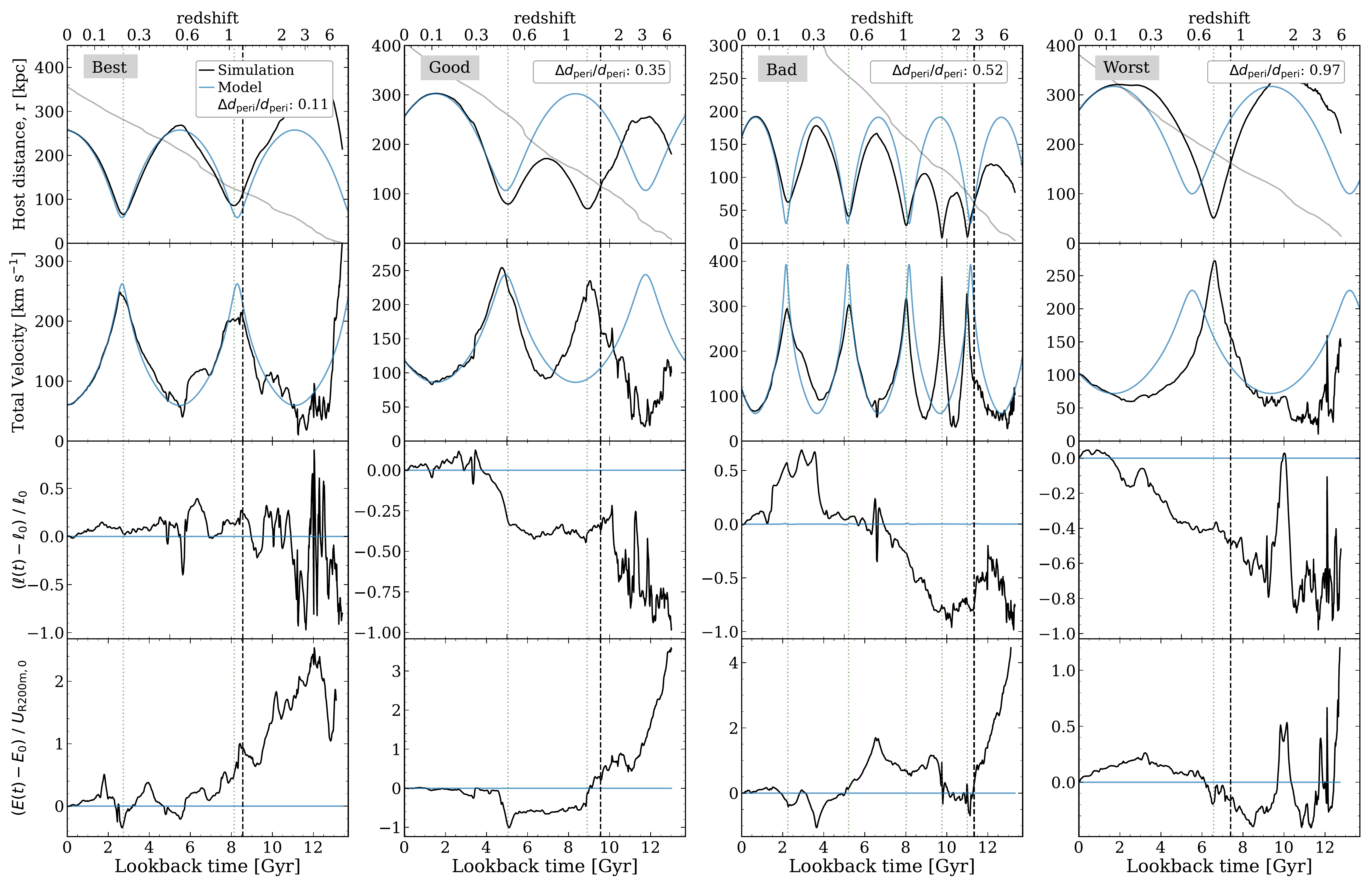}
\end{tabular}
\vspace{-2 mm}
\caption{
\textbf{Four case studies of satellite orbital histories}.
Orbital distance from the host galaxy, $r$ (top row), total velocity (second row), specific angular momentum (third row), and specific total energy (bottom row).
We show 4 satellites based on how well the most recent pericentre agrees between the simulations and orbit modeling, $(d_{\rm peri,model} - d_{\rm peri,sim}) / d_{\rm peri,sim}$, with increasing error from left to right (see legend).
Black lines show the simulations, while blue lines show model-based orbits in a static axisymmetric host potential.
Black vertical dashed lines show the first infall into the MW-mass halo in the simulation, and in the top row the grey line shows $\Rthm(t)$ of the host halo.
Green vertical dotted lines show pericentres in the simulation.
In the left case study, the orbit model and simulation agree for nearly two full orbits, while the right case study shows agreement for less than half an orbit.
Orbit modeling tends to recover the timing of the most recent pericentres better than their distances.
As the bottom rows show, specific angular momentum and total energy of the orbit are not generally conserved; see also Figures~\ref{fig:e_ell_prop} and \ref{fig:conservation}.
}
\label{fig:orbits}
\end{figure*}

We now compare orbit properties of satellites from our cosmological simulations to properties derived in an idealized, static, axisymmetric model.
As we describe in detail in Appendix~\ref{app:mass_profile}, we fit the present-day host potential, keep it fixed over time, and initialise the satellite orbits at $z = 0$ using the same 6D phase-space coordinates as in the cosmological simulations.
Thus, the orbital energy and angular momentum of satellites remain constant, and the satellites orbit periodically across $13.8 \Gyr$.

Figure~\ref{fig:orbits} shows four representative satellites: each column shows varying degrees of how well orbit modeling reproduces the most recent pericentre distance.
To quantify how well orbit modeling does in reproducing the recent pericentre distance, we measure $\Delta d_{\rm peri} / d_{\rm peri} = (d_{\rm peri, model} - d_{\rm peri, sim}) / d_{\rm peri, sim}$.
From top to bottom, we compare the host-centric distance, the total velocity, specific angular momentum, and specific energy of the orbit.

Orbit modeling agrees well with the simulations during the satellites' recent histories.
In the left two columns, orbit modeling recovers the orbits well for one half to two orbits, the third column shows agreement with the timing of the orbit for two and a half orbits but less agreement for the distance and velocity, and the right column show agreement for less than half an orbit.

Even in the left two cases, in which the model does well at reproducing the most recent pericentre distance, orbit modeling does not accurately recover previous pericentres, especially the timing of the pericentres, which continues to become more out of phase with time, likely due to the lack of dynamical friction.
The right column shows cases in which the timing of the most recent pericentre is within $\approx 0.5 \Gyr$ but the pericentre distance is off by nearly a factor of 2.

Finally, the third and bottom rows of Figure~\ref{fig:orbits} show the lack of conservation in specific angular momentum and specific energy for the satellites in the simulations.
For each satellite, we show the fractional change in $\ell$ compared to present-day, that is, $(\ell(t) - \ell_0) / \ell_0$.
Even after a satellite falls into its MW-mass halo, its angular momentum can increase or decrease $\gtrsim 50$ per cent over time.
The lack of conservation in $\ell$ is likely a combination of complex processes, including the growth of the MW-mass host, satellite-satellite interactions, mergers, and the non-symmetric potential.
In the bottom row, we calculate the fractional change in energy compared to present-day, normalised by the host virial potential energy today, that is, $(E(t) - E_0) / U_{\rm R200m, 0}$.
Similar to the results in Figures~\ref{fig:e_ell_prop}-\ref{fig:conservation}, the specific energy of a satellite orbit decreases over time, primarily because of the growth of the MW-mass host.

In the following subsections, we quantify differences in orbit properties across the entire satellite population.
It is worth noting that dynamical friction acts more efficiently at higher masses to rob the satellites of their orbital energy and cause them to merge away \citep[for example][]{BoylanKolchin08}.
We remind the reader that when interpreting the plots, we do not include a model for dynamical friction.

\subsubsection{Orbital distance} %
\label{sec:orbit_comp}         %

\begin{figure}
\centering
\begin{tabular}{c}
\includegraphics[width = 0.95 \linewidth]{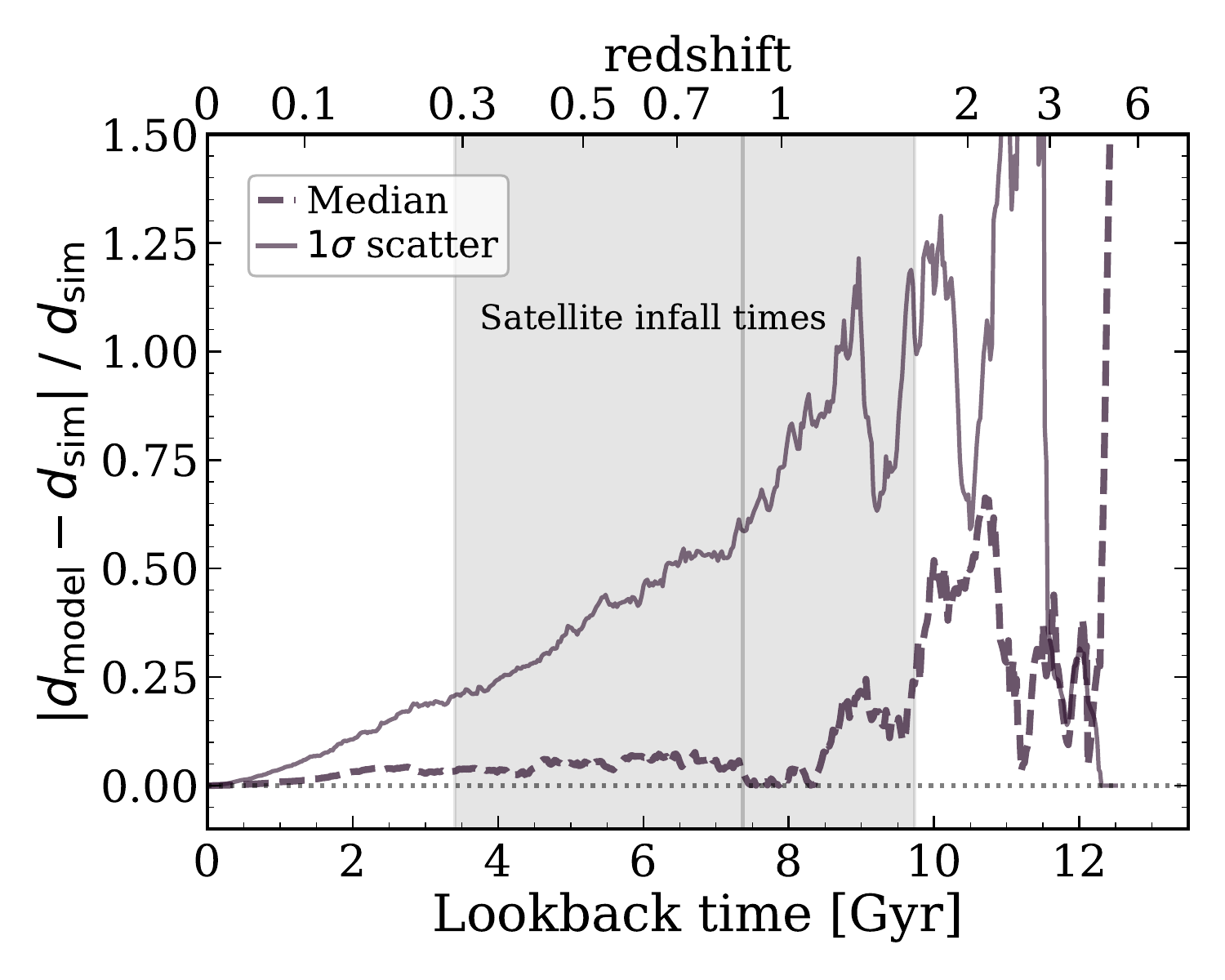}
\end{tabular}
\vspace{-2 mm}
\caption{
The fractional difference between orbit modeling and the simulations for the host-centric  distance versus lookback time.
The dashed line shows the absolute value of the median across all satellites and the solid line shows the $1 \sigma$ scatter of the sample.
We choose to show the absolute value of the fractional difference to keep all values positive for visual clarity, however, the median is negative for lookback times $\lesssim7.5\Gyr$, and positive for earlier times.
Similar to Figure~\ref{fig:conservation}, we only include the instantaneous population of satellites for a given lookback time.
The median reaches 25 per cent at $9.7 \Gyr$ ago, but the $1 \sigma$ scatter reaches 25 per cent already at $4 \Gyr$ ago.
}
\label{fig:model_comp}
\end{figure}

First, Figure~\ref{fig:model_comp} shows the absolute fractional difference in host-centric distance from the simulations versus from orbit modeling, versus lookback time.
We show the absolute value of the median to keep all values positive and visual clarity, however, the median is negative for lookback times of $\tlb\lesssim7.5\Gyr$, and positive for $\tlb\gtrsim7.5\Gyr$.
Over the last $8.5 \Gyr$, the median fractional difference is relatively constant at $\lesssim 10$ per cent.
Before this, the median then increases to 25 and 50 per cent, around $9.7$ and $10 \Gyr$ ago.
Prior to $\approx 11.7 \Gyr$ ago, less than 1 per cent of these satellites today were still satellites.
The $1 \sigma$ scatter reaches a 25, 50, and 100 per cent fractional difference at $4, 6.3,$ and $8.6 \Gyr$ ago.


\subsubsection{Virial infall time} %
\label{sec:infall}                 %

\begin{figure*}
\centering
\begin{tabular}{c}
\includegraphics[width = 0.95 \linewidth]{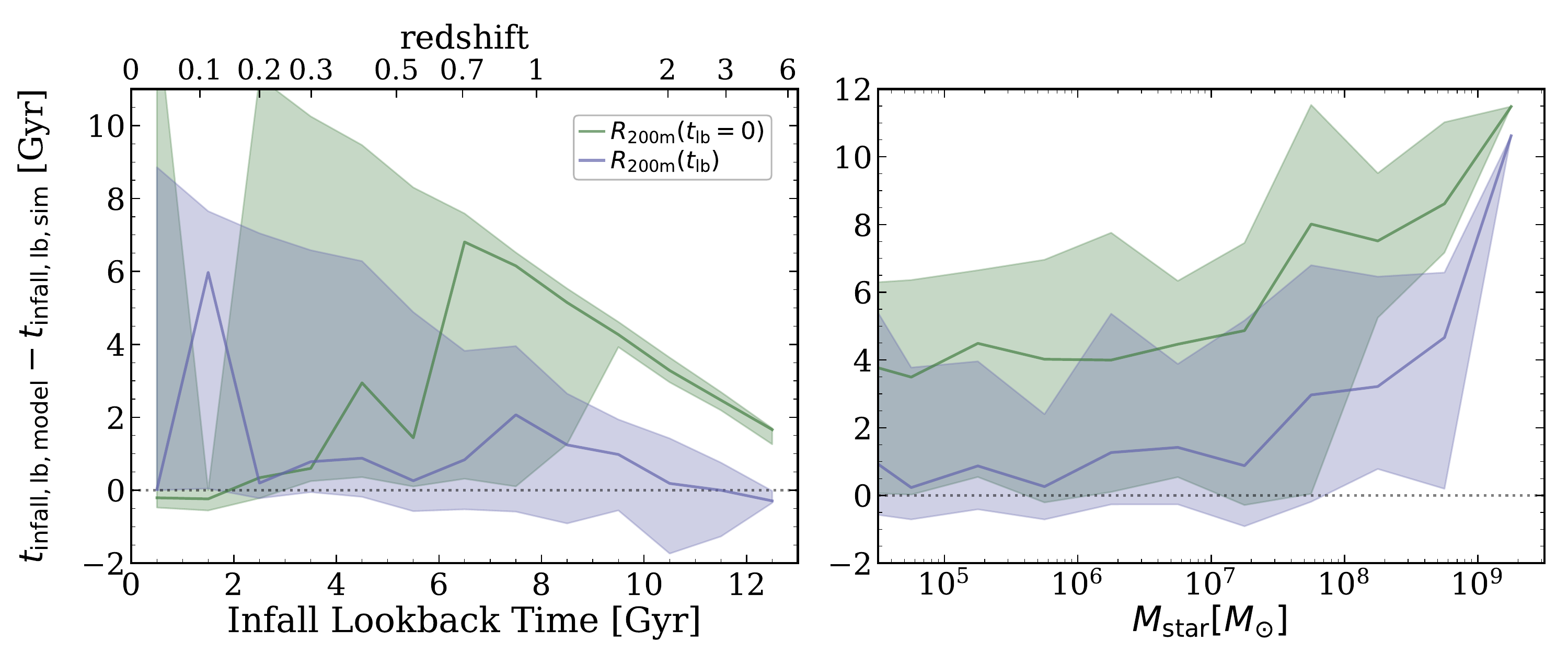}
\end{tabular}
\vspace{-2 mm}
\caption{
The difference in infall time, from orbit modeling versus the simulations, as a function of infall time in the simulations (left) and satellite $\Mstar$ (right).
For orbit modeling, we measure infall time two ways: using a non-evolving host halo radius, $\Rthm(\tlb = 0)$ (green) and using $\Rthm(\tlb)$ from the simulations (purple).
Solid lines show the median and shaded regions show the 68th percentile ranges across our satellites.
\textbf{Left}: The median infall time is generally accurate for orbit modeling if using an accurate $\Rthm(\tlb)$, with a median offset of $\lesssim 2 \Gyr$ and a typical scatter of $2.4 \Gyr$.
The spike at $1.5 \Gyr$ comes from orbit modeling overpredicting the infall times for most satellites by $\gtrsim 5 \Gyr$, given errors in modeling recent apocentre distances.
By contrast, orbit modeling using fixed $\Rthm(\tlb = 0)$ works well for recently infalling satellites but vastly overestimates the lookback time to infall for earlier-infalling satellites, with an average $1 \sigma$ scatter of $2.8 \Gyr$.
\textbf{Right}: The median and 68th percentile for both infall time metrics show similar trends with satellite mass, offset by $\approx 3 \Gyr$.
The typical $1 \sigma$ scatters range from roughly $3 \Gyr$ and $2.5 \Gyr$ when using either the fixed $\Rthm(\tlb = 0)$ or accurate $\Rthm(\tlb)$, respectively.
Orbit modeling fails most significantly for satellites with $\Mstar \gtrsim 10^7 \Msun$, because dynamical friction has shrunk their orbits.
}
\label{fig:infall_comp}
\end{figure*}

Many studies focus on when low-mass galaxies first become satellites, and their properties, such as mass, during infall \citep[for instance][]{BoylanKolchin11, Wetzel15, Patel17}.
We investigate two ways of calculating the time of first infall into the host halo within orbit modeling, $t_{\rm lb, infall, model}$.
First, when a satellite first crossed within the MW-mass halo, accounting for the growth of its $\Rthm$ over time.
Second, we record when a satellite's orbit first crossed within $\Rthm$ at $z = 0$.
Nearly 60 per cent of all satellites in orbit modeling always have orbited within $\Rthm(\tlb = 0)$, so we simply define these infall times to be $13.8 \Gyr$ ago.
We take the difference of each $t_{\rm lb, infall, model}$ to the infall time in the simulations, $t_{\rm lb, infall, sim}$ (calculated with an evolving $\Rthm$).
Figure~\ref{fig:infall_comp} shows these differences, versus the infall times in the simulations (left) and versus satellite $\Mstar$ (right).

In the left panel, using an evolving $\Rthm(\tlb)$, the median difference between orbit modeling and the simulations is generally within $\approx 2 \Gyr$.
The peak in the curve at $1.5 \Gyr$ is driven by 11 of the 20 satellites in this particular bin in infall time, where orbit modeling predicts that these galaxies fell in $\gtrsim 5 \Gyr$ earlier than in the simulations.
Similarly, a slightly smaller peak at $7.5 \Gyr$ is caused by the model over-predicting $t_{\rm lb, infall, model}$ for nearly half of the satellites by $\gtrsim 2 \Gyr$.

The 68th percentile range is largest for the most recently infalling satellites and decreases with increasing lookback time.
Orbit modeling generally overestimates the infall time compared to the simulations, because the model orbits are periodic and more likely to cross the virial radius at earlier times.
The model overpredicts the infall time for roughly 65 per cent of all satellites.
Even when accounting for the evolving $\Rthm$, the $1 \sigma$ scatter spans $\gtrsim 1 \Gyr$, which highlights the large uncertainty in orbit modeling.

When using a fixed $\Rthm(\tlb = 0)$, the median shows relatively good agreement for satellites that fell in within the last $\approx 4 \Gyr$.
Beyond $4 \Gyr$ ago, the median difference in infall time increases until $\approx 7 \Gyr$ ago, where it decreases again.
The orbits of these satellites were generally within $\Rthm(\tlb = 0)$ at all times, so the difference between orbit modeling and simulations follows the relation $13.8 \Gyr - t_{\rm lb, infall, sim}$.
Of the subset of satellites that fell in between $1 - 2 \Gyr$ ago, only 2 of the 17 satellites were always within $\Rthm$, which is why the 68th percentile dips down close to 0.
However, the associated uncertainties in this method of calculating infall time are much worse, and the $1\sigma$ scatter reaches as large as $\approx 5.7 \Gyr$.

Versus $\Mstar$, both infall metrics follow the same general trends: better agreement for satellites with $\Mstar < 10^7 \Msun$ and larger offsets for higher-mass satellites.
The offset between the medians, and 68th percentiles, is roughly $3 - 4 \Gyr$ between the two infall metrics.
The values associated with the fixed $\Rthm(\tlb = 0)$ method skew to larger values, because many of the model orbits always orbited within this distance.
The $1 \sigma$ scatters each span roughly $2.5 - 3 \Gyr$, so one cannot accurately determine a satellite galaxy's infall time at any given mass to within $\approx 2.5 \Gyr$.

\subsubsection{Pericentre properties} %
\label{sec:peri}                      %

We next investigate various properties associated with pericentric passages relative to the host galaxy, when the tidal acceleration and ram pressure from the host CGM tend to be strongest.

Most works assume that satellite orbits only \textit{shrink} over time, because of dynamical friction and the time-dependent host potential \citep[for example][]{Weinberg86, Taylor01, Amorisco17}, which implies that the most recent pericentre should be the smallest experienced.
However, as we showed in \citet{Santistevan23}, for 67 per cent of our satellites with $N_{\rm peri} \geq 2$ the most recent pericentre is not the smallest, because many satellite orbits have grown in pericentre distances over time.
\citet{Patel20} also saw cases in which the most recent pericentre was not the smallest, and suggest that the presence of a massive satellite alone can cause this effect.
Therefore, we present trends for the most recent and the minimum pericentres.

\begin{figure*}
\centering
\begin{tabular}{c}
\includegraphics[width = 0.95 \linewidth]{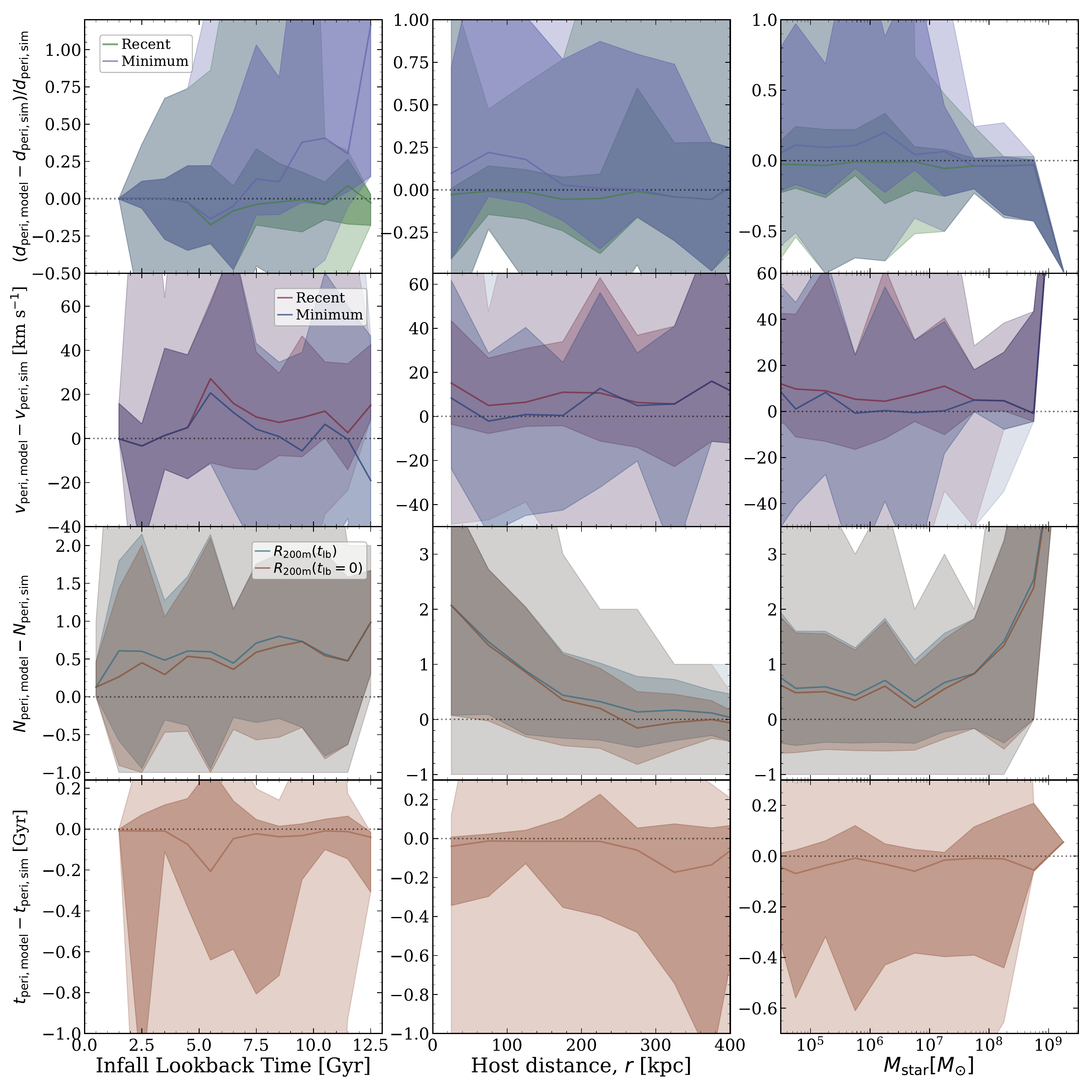}
\end{tabular}
\vspace{-2 mm}
\caption{
Comparing various properties of orbital pericentres, between the simulations and orbit modeling, for surviving satellites versus their lookback time of infall into the MW-mass halo (left), present-day distance from the MW-mass host, $r$ (middle), and satellite $M_{\rm star}$ (right).
Solid lines show the median, and the shaded regions show the 68th and 95th percentiles, across all satellites.
\textbf{Top row}: Fractional difference between pericentre distances, $(d_{\rm peri,model} - d_{\rm peri,sim}) / d_{\rm peri,sim}$, for both the most recent and the minimum pericentre.
Orbit modeling predicts larger minimum pericentres, as high as 100 per cent in the median, for satellites that fell in $\gtrsim 12 \Gyr$ ago (left), and within 25 per cent for satellites at small $r$ (middle), and for satellites $\lesssim 10^7 \Msun$ (right).
\textbf{Second row}: Difference between the total velocity at pericentre, $v_{\rm peri,model} - v_{\rm model,sim}$, for both the most recent and the minimum pericentre.
Orbit modeling generally overpredicts both pericentre velocities by $\approx 10 - 20 \kmsi$.
\textbf{Third row}: Difference between the mean number of pericentric passages about the MW-mass host, since first crossing the growing host $\Rthm(\tlb)$ and since first crossing $\Rthm(\tlb = 0)$.
This difference slightly increases from $\approx 0$ to $1$ with $t_{\rm lb}$ (left) and decreases from 2 to 0 with $r$ (middle), because satellites at small distance typically fell in earlier, which means that they orbited longer in the model.
\textbf{Fourth row}: Difference between the lookback time of the most recent pericentre, $t_{\rm peri, model} - t_{\rm peri, sim}$, which shows weak trends with any properties.
Although the median trends across the satellite population agree well in most cases, the substantial scatter in all panels implies \textit{significant} uncertainty for a given satellite's orbit history.
}
\label{fig:peri_props}
\end{figure*}

Figure~\ref{fig:peri_props} compares pericentre distances, $d_{\rm peri}$, the velocity at pericentre, $v$, the number of pericentric passages, $N_{\rm peri}$, and the timing of pericentres, $t_{\rm peri}$, versus the lookback time of infall into the MW-mass halo, present-day distance, $r$, and satellite $\Mstar$.
For each pericentre property, we show the difference between orbit models and simulations, except for pericentre distance, for which we compare the \textit{fractional} difference.

Figure~\ref{fig:peri_props} (top row) shows the trends for pericentre distance, for both the most recent, $\dperirec$, and the minimum pericentre, $\dperimin$.
For a given satellite, all of its pericentre distances are the same in our (static) orbit models.
With respect to satellite infall time (left panel), the model recovers the median $\dperirec$ well, but this is across the entire population of satellites, not for a given satellite.
The median fractional difference is within $\approx 20$ per cent, and the average $1 \sigma$ scatter is roughly $19$ per cent, smaller than many other properties we present here.
Thus, although the median pericentre distance across the population looks reasonable, the prediction for any particular satellite is uncertain by $\approx 20$ per cent.
Orbit modeling recovers the median $\dperimin$ well for satellites that fell in $\lesssim 5 \Gyr$ ago, because all but one of these satellites experienced only one pericentre, so the minimum is the most recent.
For satellites that fell in $\gtrsim 5 \Gyr$ ago, the median fractional offset in $\dperimin$ diverges from that of $\dperirec$.
Roughly 60 per cent of satellites that fell in $\gtrsim 5 \Gyr$ ago experienced multiple pericentres, and because the orbit models only predict a single $d_{\rm peri}$ for a given satellite, these positive values suggest that the satellites in the simulations orbit at closer distances than in our static model.
The $1 \sigma$ scatter for $\dperimin$ reaches 100 per cent around $9.5 \Gyr$ ago, thus one cannot accurately predict the minimum pericentre to within $\gtrsim 2$ for satellites that fell in earlier than this.

In the middle panel, median $\dperimin$ and $\dperirec$ between the orbit models and simulations show general consistency across all distances
The fractional difference in $\dperirec$ is $\lesssim 5$ per cent, and $\lesssim 25$ per cent for $\dperimin$.
As we will discuss below, $\approx 95$ per cent of satellites that currently orbit beyond $300 \kpc$ completed only one pericentre, so the median and 68th percentiles are the same between both $\dperimin$ and $\dperirec$.
Conversely, 2/3 of satellites currently within $300 \kpc$ have $\Nperi \geq 2$ and, similar to the left panel, the median and scatter for $\dperimin$ increases to positive values, indicating that orbit modeling overpredicts $\dperimin$ for these satellites.
For the satellites within $300 \kpc$, nearly 85 per cent fell into their MW-mass hosts over $5 \Gyr$ ago.
As with the left panel, the range in the $1 \sigma$ scatter is larger for $\dperimin$ than in $\dperirec$, with average values of 55 per cent and 24 per cent, respectively.

Finally, the median fractional difference in both pericentre metrics shows no dependence on stellar mass for $\Mstar < 10^{8.75} \Msun$ (right panel).
Lower-mass satellites typically fell in earlier, orbit at closer distances, and completed more pericentres, so their minimum and most recent pericentres are more likely to diverge.
Only one satellite in our sample has $\Mstar > 10^9 \Msun$.

Figure~\ref{fig:peri_props} (second row) shows trends in the total velocity of pericentre, $v_{\rm peri}$, and similar to $d_{\rm peri}$, we show trends in both the minimum, $\vperimin$, and most recent, $\vperirec$.
Again, satellites at small infall time, large $r$, and high $\Mstar$, only experienced one pericentre, so the median trends in both $\vperimin$ and $\vperirec$ are the same.
The model recovers both the median $\vperimin$ and $\vperirec$ to within $\approx 30 \kmsi$ across all infall times and $r$, and nearly all $\Mstar$.
Although the model recovers the median $\dperirec$ well, it over-estimates $\vperirec$ in all panels presumably because of the lack of dynamical friction and gravitational perturbations from other satellites.

Figure~\ref{fig:peri_props} (third row) compares the number of pericentric passages a satellite experienced, $\Nperi$.
For the model orbits, we only count the number of pericentric passages a satellite experienced since first infall.
We count $\Nperi$ from the two infall metrics in Section~\ref{sec:infall}: since infall into the MW-mass halo accounting for an evolving $\Rthm(\tlb)$, and since infall while keeping a fixed $\Rthm$ $\Rthm(\tlb = 0)$.
We show the \textit{mean} and standard deviation for $\Nperi$, because it is an integer for a given satellite.

These results for $\Nperi$ are not particularly sensitive to the two ways in which we calculate first infall.
In the left panel, the mean difference in $\Nperi$ slightly increases with infall time, because orbits in the models are periodic, so longer integration times lead to larger $\Nperi$.
The middle panel shows trends versus host distance, $r$: for satellites at $\lesssim 250 \kpc$, orbit modeling overpredicts $\Nperi$, because these satellites typically fell into their MW-mass hosts earlier than satellites at larger $r$.
Finally, the mean difference in $\Nperi$ is generally flat at $\Mstar < 10^{7.5} \Msun$, but the difference increases for more massive satellites, given the lack of dynamical friction in the orbit models.


Finally, Figure~\ref{fig:peri_props} (bottom row) compares the timing of just the most recent pericentre, $\tperi$.
The median difference is $\lesssim 0.2\Gyr$ across all three panels.
The median is also consistently negative, indicating that orbit modeling predicts more recent pericentres, likely because in the orbit models the MW-mass host does not reduce in mass going back in time.


\begin{figure*}
\centering
\begin{tabular}{c}
\includegraphics[width = 0.95 \linewidth]{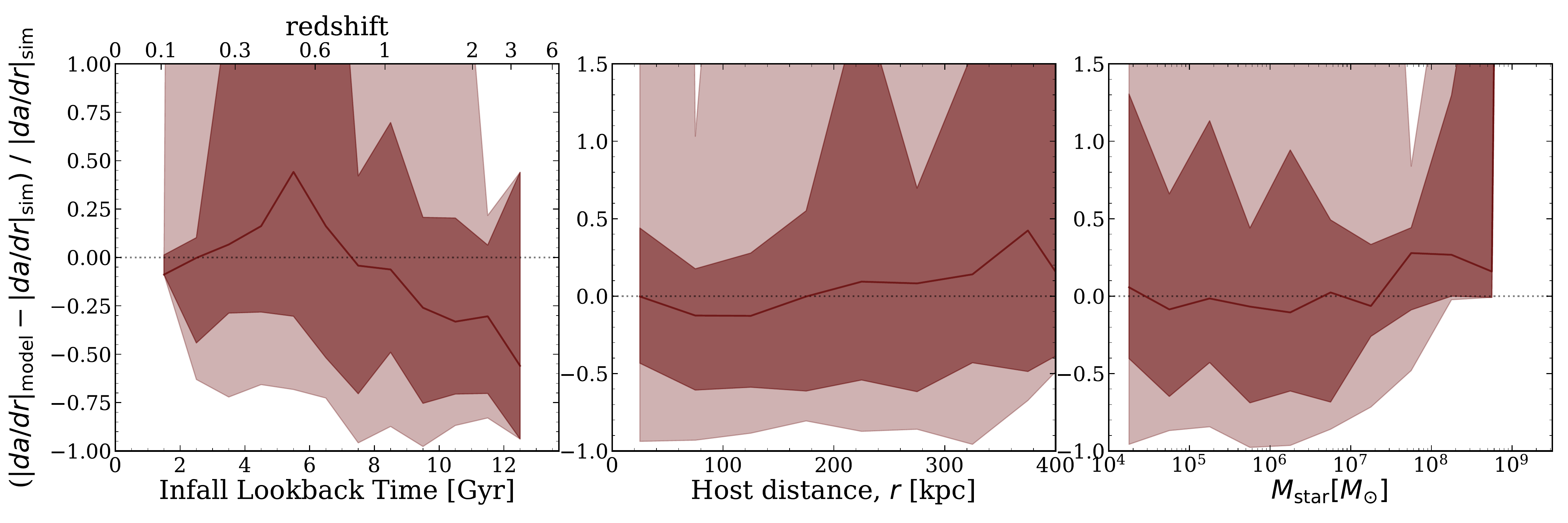}
\end{tabular}
\vspace{-2 mm}
\caption{
Comparing the maximum tidal acceleration, $\dadr$, from the MW-mass host that satellites experienced via orbit modeling versus in the simulations, as a function of lookback time of infall into the MW-mass halo (left), distance from the host, $r$ (middle), and satellite $\Mstar$ (right).
Solid lines show the median and the dark and light shaded regions show the 68th and 95th percentiles across all satellites.
These trends  mirror those for the minimum pericentre distance, $\dperimin$, in Figure~\ref{fig:peri_props} (top).
Because this orbit modeling does not account for the growth of the MW-mass host and the orbits are periodic, it increasingly over-predicts the pericentre distance, and underpredicts the tidal acceleration, with increasing $\MWinfall \gtrsim 5 \Gyr$.
The dependence on $r$ and $\Mstar$ are weak.
However, the $1 \sigma$ scatters span more than 50 per cent in each of the panels here, and up to a factor of 2, highlighting the large uncertainties.
}
\label{fig:dadr_comp}
\end{figure*}

Typically maximized near a pericentric passage, satellites feel a tidal acceleration from the host, which strips their mass.
We calculate the acceleration to be $a = G M(< r) / r^2$, where $M(< r)$ is the total enclosed mass of the host within a distance $r$.
We then compute the derivative with respect to $r$ and save the maximum $\dadr$ that a satellite experienced after first infall.

Figure~\ref{fig:dadr_comp} compares the maximum $\dadr$ experienced between the simulation and model.
Satellites that fell in $\MWinfall = 2.5 - 7 \Gyr$ ago typically have larger minimum pericentres in the simulations than in orbit modeling, so the model overpredicts $\dadr$ for these satellites by up to 45 per cent.
Conversely, satellites that fell in $\MWinfall \gtrsim 7 \Gyr$ ago show larger minimum pericentres in the orbit models, so underpredicts $\dadr$ for the earliest satellites by up to 55 per cent.
Because the simulations and orbit models agree in $\dperimin$ for satellites that fell in $\MWinfall < 2.5 \Gyr$ ago, the median fractional difference is near zero.

Figure~\ref{fig:dadr_comp} (middle and right) shows that $\dadr$ has little to no dependence on present-day satellite distance or $\Mstar$.
Although the median fractional difference is close to 0 in both panels, the $1 \sigma$ scatter increases from $0.39 - 2$ with $r$, and the mean scatter versus $\Mstar$ is 73 per cent.
In all three panels, the $2 \sigma$ scatter spans $100$ per cent or more, so while the median $\dadr$ across the population from orbit modeling is relatively accurate, for any given satellite, orbit modeling over- or underpredicts $\dadr$ typically by a factor of $\approx 2$.


Finally, Appendix~\ref{app:peri_comp} compares both the timing and distance of pericentres between orbit modeling and the simulations at each previous lookback pericentric event.
The bias in both the distance and timing of pericentres increases with increasing lookback pericentre events to roughly 20 per cent in distance, and $\approx 1.5 \Gyr$ in time.
The uncertainty in these measurements increases up to the 4th most recent pericentre to $\approx 50$ per cent in distance and $\approx 1 \Gyr$ in time.
Beyond this, $\lesssim 8$ per cent of satellites experienced 5 pericentres or more.

In summary, we compared various pericentre properties for both the minimum and most recent pericentre events.
Across the full sample, the median fractional difference, or bias across the population, for the minimum and most recent pericentre distances are $2.5 - 6.6$ per cent.
The bias in the pericentre velocity is within $\approx 20 \kmsi$, within $0.2 \Gyr$ for the timing of the most recent pericentre, and $< 2$ for the number of pericentric events.
Finally, the bias in the maximum tidal acceleration is typically 10's of per cent across infall time, $r$, and $\Mstar$.
Just as importantly, the typical $1 \sigma$ scatter, which represents the uncertainty for a given satellite, is significant at $\approx 10 - 70$ per cent.

\subsubsection{Apocentre, orbital period, and eccentricity} %
\label{sec:other_orbit}                                     %

\begin{figure*}
\centering
\begin{tabular}{c}
\includegraphics[width = 0.95 \linewidth]{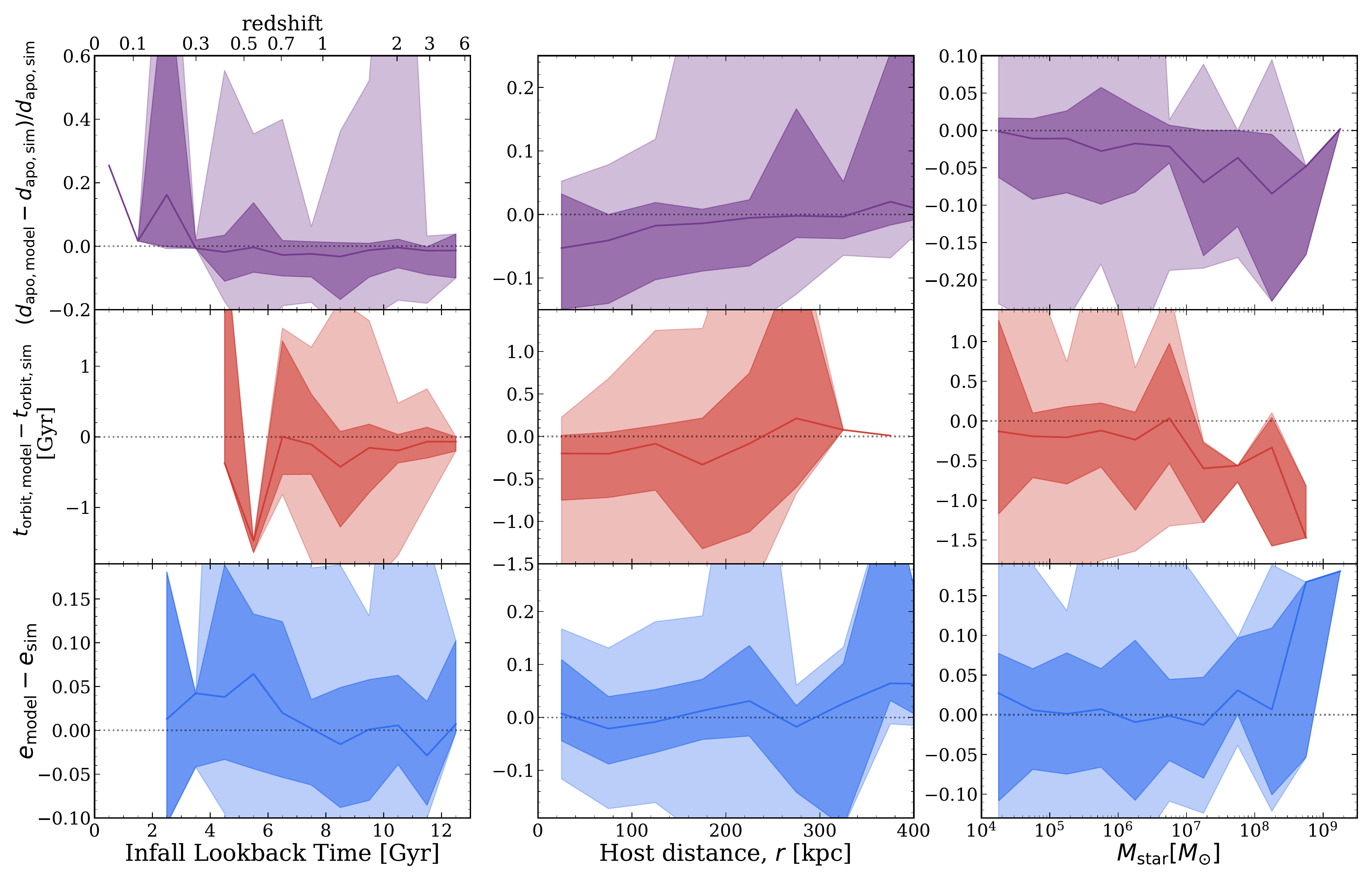}
\end{tabular}
\vspace{-2 mm}
\caption{
Comparing the most recent apocentre distance, $d_{\rm apo}$, orbital period, $t_{\rm orbit}$, and orbital eccentricity, $e$, from orbit modeling versus the simulations, as a function of lookback time of infall in the MW-mass halo (left), current distance from MW-mass host, $r$ (middle), and satellite $\Mstar$ (right).
Solid lines show the median and the dark and light shaded regions show the 68th and 95th percentiles across all satellites.
\textbf{Top row}: The most recent apocentre distance.
Versus infall time, the median is roughly constant at $-2$ per cent for satellites that fell in $\gtrsim 3.5 \Gyr$ ago.
The model recovers the median apocentre distance to within $\pm 5$ per cent versus $r$, but the median decreases slightly with $\Mstar$ from $\approx 0$ to 8 per cent at $\Mstar = 10^{8.25} \Msun$.
Although the medians in each panel may only be within a few per cent, the $1 \sigma$ scatters span $\approx5-10$ per cent or more, highlighting the uncertainty for a given satellite.
\textbf{Middle row}: Difference between the most recent orbital time, $t_{\rm orbit}$, defined as the difference in time between the two most recent pericentres.
Because orbit modeling generally under-predicts the recent pericentre lookback times, and because the orbits are periodic, the median difference in $t_{\rm orbit}$ is slightly negative across all panels, and even as low as $0.5 - 1.5 \Gyr$ for satellites with $\Mstar \gtrsim 10^{7.25} \Msun$.
\textbf{Bottom row}: Difference between most recent orbital eccentricity, $e = (d_{\rm apo} - d_{\rm peri}) / (d_{\rm apo} + d_{\rm peri})$.
The difference in $e$ varies by at most $0.06$ versus $\MWinfall$ and $r$, but satellites with $\Mstar > 10^{8.5} \Msun$ have differences $> 0.15$.
In general, orbit modeling recovers the median properties here within $\approx 7$ per cent (see Table~\ref{tab:summary_stats}), though with significant scatter.
The $1 \sigma$ scatters span $8-15$ per cent, likely because these properties all depend on pericentre and apocentre events that occur in the recent past.
}
\label{fig:apo_n_friends}
\end{figure*}

The apocentre measure how far a satellite orbits from its host, and an orbit spends most of its time near apocentre.
Figure~\ref{fig:apo_n_friends} (top) compares trends in the most recent apocentre distance, $\daporec$.
We only measure an apocentre that occurs after infall into the MW-mass halo.
About 68 per cent of our satellites experienced an apocentre; the rest are on first infall.

Versus infall time (top left),
8 satellites fell into the host between $\MWinfall = 2 - 3 \Gyr$ ago, and orbit modeling generally overpredicts $\daporec$ by $\gtrsim 75$ per cent for half of them.
The fractional difference in apocentre distance is smaller for earlier-infalling satellites, and the median is $\approx 0.015$ with a mean $1 \sigma$ scatter of $0.06$.
Figure~\ref{fig:apo_n_friends} (top middle) shows little dependence with $r$.
Satellites that currently orbit at smaller distances generally fell into the MW-mass host earlier, so orbit modeling somewhat underpredicts $\daporec$ for satellites within $\lesssim 250 \kpc$, similar to how it underpredicted $\daporec$ at large $\MWinfall$ (top left).
Overall, the mean $1 \sigma$ scatter is $\approx 0.08$.
Finally, the median fractional difference in apocentre distance decreases weakly with $\Mstar$ (top right).
Lower-mass satellites typically fell into their MW-mass halo earlier, and they have smaller fractional differences.

Figure~\ref{fig:apo_n_friends} (middle row) shows trends in the most recent orbital period, $t_{\rm orbit}$.
We define the orbit period as the time difference between the two most recent pericentric passages,
and we find nearly identical results using the times between apocentres.
47 per cent of the satellites in our sample experienced 2 or more pericentres in both the simulations and orbit modeling.

In the left panel, satellites that fell in $\MWinfall < 4.5 \Gyr$ ago did not have enough time to undergo 2 pericentres.
For earlier infalling satellites, the median difference in $t_{\rm orbit}$ varies by as much as $-0.4 \Gyr$, but the mean across all infall times is $-0.15$.
The difference in $t_{\rm orbit}$ is negligible versus $r$.
Orbit modeling does not account for dynamical friction, therefore, for satellites with $\Nperi \geq 2$, if orbit modeling underpredicts $\dperirec$ compared to the simulations, it suggests more bound orbits and smaller $t_{\rm orbit}$ values, which is what we see for these satellites with $\Mstar \gtrsim 10^{6.5} \Msun$.

Finally, Figure~\ref{fig:apo_n_friends} (bottom row) compares the orbital eccentricities, $e$.
Figure~\ref{fig:apo_n_friends} (bottom left) shows that, for satellites that fell in $\MWinfall \lesssim 4 \Gyr$ ago, orbit modeling recovers $\dperirec$ well (Figure~\ref{fig:peri_props}, top left) and overpredicts $\daporec$.
Similarly, although orbit modeling recovers $\daporec$ well for satellites up to $\MWinfall = 7.5 \Gyr$ ago, it also underpredicts $\dperirec$, which drives $e$ to be higher in orbit modeling.
Orbit models and simulations show similar results for earlier-infalling satellites.
The median difference in $e$ is flat with $r$ and $\Mstar < 10^{8.25} \Msun$.

Again, we compute all results in this subsection based on the most recent pericentre and apocentre, but as Figure~\ref{fig:peri_props} showed, orbit modeling performs worse for earlier properties of an orbit, so comparison of orbital period and eccentricity at earlier stages of these orbits would show even larger disagreements.

\subsubsection{Recoverability of orbit properties} %
\label{sec:stats}                                  %

We compared 15 properties of satellite orbits in our cosmological simulations against orbit models using static axisymmetric potentials that we fit near-exactly to our hosts at $z = 0$.
Table~\ref{tab:summary_stats} lists the properties that we tested, as well as the median offsets and $1 \sigma$ and $2 \sigma$ scatters across our sample of satellites.
We compare both the raw difference of a given orbit property, $X$, defined as $X_{\rm model} - X_{\rm sim}$, as well as the fractional difference, $(X_{\rm model} - X_{\rm sim}) / X_{\rm sim}$.
Additionally, we show the fractional change in orbital specific energy since infall relative to the MW-mass potential, $(E_0 - E_{\rm inf}) / U_{\rm 200m, 0}$ and the fractional change in orbital specific angular momentum relative to today, $(\ell_0 - \ell_{\rm inf}) / \ell_0$. The orbit models conserve these quantities by definition.

We quantify the goodness of the orbit models in terms of their `bias' (accuracy) and `uncertainty' (precision) in the right-most columns of Table~\ref{tab:summary_stats}.
The `bias' describes how well orbit modeling accurately recovers the median orbital property across the satellite population: we define a property to be minimally, moderately, or highly biased if the median fractional offset of the satellite population between orbit modeling and the simulations is $< 10$ per cent, $10 - 25$ per cent, or $> 25$ per cent, respectively.
However, even in cases where this bias is small (accuracy is high), orbit modeling can have severe limitations if it cannot model the history of a given satellite to good precision.
Thus, we also quantify the `uncertainty' via how large the scatter in this difference between orbit models and simulations is across the satellite population.
We define a property to be minimally, moderately, or highly uncertain if the $1 \sigma$ scatter is $< 25$ per cent, $25 - 50$ per cent, or $> 50$ per cent.
Because bias is more problematic (systematic) than uncertainty, we impose stricter criteria for it.

\begin{table*}
\caption{
Comparing the results of orbit modeling in a static axisymmetric host potential against cosmological baryonic simulations.
Column list: property name; variable; median offset, $1 \sigma$ scatter, $2 \sigma$ scatter.
The left columns compare the raw difference, $X_{\rm model} - X_{\rm sim}$, while the middle columns compare the fractional difference, ($X_{\rm model} - X_{\rm sim}$) / $X_{\rm sim}$.
Additionally, we describe the strength of the bias (median fractional offset) and uncertainty ($1 \sigma$ scatter of the fractional offset) associated with each property.
In the bottom two rows, we show the difference in the orbital energy between present day and the infall time, normalised by the host halo potential energy at present-day, $(E_0 - E_{\rm inf}) / U_{\rm 200m,0}$, and the fractional change in the orbital specific angular momentum relative to present-day, $(\ell_0 - \ell_{\rm inf}) / \ell_0$.
Given that energy and angular momentum are always conserved in the model, we place them below a horizontal line to distinguish them.
}
\begin{tabular}{|l c |@{\vline}| c@{\hspace{-0.1 ex}}c@{\hspace{-0.1 ex}}c |@{\vline}| c@{\hspace{-0.1 ex}}c@{\hspace{-0.1 ex}}c |@{\vline}| c|c|}
\hline
\hline
& & & \underline{Raw Difference} & & & \underline{Fractional Difference} & \\
Orbital property  & Variable   & Median  & $1 \sigma$ & $2 \sigma$ & Median & $1 \sigma$ & $2 \sigma$ & Bias & Uncertainty   \\
&            & offset  & scatter    & scatter    & offset & scatter    & scatter    &      &   \\
\hline
Recent pericentre & $d_{\rm peri, rec}$  [$\kpc$]     & -1.26  & 12.1 & 68.8 & -0.025 & 0.21 & 1.19 & Min  & Min   \\
distance          &                                   &        &      &      &        &      &      &      &       \\
Min pericentre & $d_{\rm peri, min}$  [$\kpc$]    &  3.23  & 18.5 & 80.1 &  0.066 & 0.53 & 3.56 & Min  & High  \\
distance &                                   &        &      &      &        &      &      &      &       \\
Lookback time of & $t_{\rm peri, rec}$  [$\Gyr$]      & -0.03  & 0.25 & 1.72 & -0.028 & 0.09 & 0.64 & Min  & Min   \\
recent pericentre          &                                 &        &      &      &        &      &      &      &       \\
Number of pericentres  & $N_{\rm peri}$               &  0.63  & 1.14 & 2.28 &  0.32  & 0.72 & 1.44 & High & High  \\
within $R_{\rm 200m}(t)$ &                                   &        &      &      &        &      &      &      &       \\
Number of pericentres & $N_{\rm peri, fixed}$          &  0.53  & 1.18 & 2.36 &  0.24  & 0.74 & 1.48 & Mod  & High  \\
within $R_{\rm 200m,0}$              &        &      &      &        &      &      &      &       \\
Velocity at & $v_{\rm peri, rec}$  [$\kmsi$]  &  7.69  & 26.8 & 106  &  0.030 & 0.10 & 0.38 & Min  & Min   \\
recent pericentre          &                                 &        &      &      &        &      &      &      &       \\
Velocity at & $v_{\rm peri, min}$  [$\kmsi$]  &  3.10  & 43.4 & 132  &  0.012 & 0.17 & 0.43 & Min  & Min   \\
min pericentre &                                 &        &      &      &        &      &      &      &       \\
Recent apocentre & $d_{\rm apo, rec}$  [$\kpc$]      & -2.75  & 12.1 & 97.3 & -0.013 & 0.06 & 0.38 & Min  & Min   \\
distance          &                                   &        &      &      &        &      &      &      &       \\
Lookback time of  & $t_{\rm inf}$  [$\Gyr$]           &  0.57  & 2.33 & 5.53 &  0.09  & 0.41 & 2.50 & Min  & Mod   \\
infall within $R_{\rm 200m}(t)$ &                                     &        &      &      &        &      &      &      &       \\
Lookback time of & $t_{\rm inf, fixed}$  [$\Gyr$]     &  4.17  & 3.55 & 8.80 &  0.44  & 0.55 & 2.92 & High & High  \\
infall within $R_{\rm 200m,0}$          &                 &        &      &      &        &      &      &      &       \\
Recent eccentricity  & $e_{\rm rec}$                         &  0.003 & 0.07 & 0.21 &  0.005 & 0.15 & 0.54 & Min  & Min   \\
          &                               &        &      &      &        &      &      &      &       \\
Recent period & $T_{\rm rec}$  [$\Gyr$]         & -0.19  & 0.49 & 1.84 & -0.071 & 0.13 & 0.41 & Min  & Min   \\
          &                                     &        &      &      &        &      &      &      &       \\
Max tidal acceleration & $\dadr$  [$\Gyr^{-2}$]              & -0.07  & 15.7 & 353  & -0.017 & 0.66 & 4.20 & Min  & High  \\
&                                &        &      &      &        &      &      &      &       \\
\hline
Energy change  & $(E_0-E_{\rm inf})/U_{\rm 200m,0}$   &  -     &  -   & -    & -0.54  & 0.82 & 1.83 & High & High  \\
since infall          &                               &        &      &      &        &      &      &      &       \\
Angular momentum  & $(\ell_0-\ell_{\rm inf})/\ell_0$  &  -     &  -   & -    & -0.015 & 0.42 & 1.68 & Min  & Mod  \\
change since infall             &                     &        &      &      &        &      &      &      &       \\
\hline
\hline
\end{tabular}
\label{tab:summary_stats}
\end{table*}

\begin{figure*}
\centering
\begin{tabular}{c}
\includegraphics[width = 0.95 \linewidth]{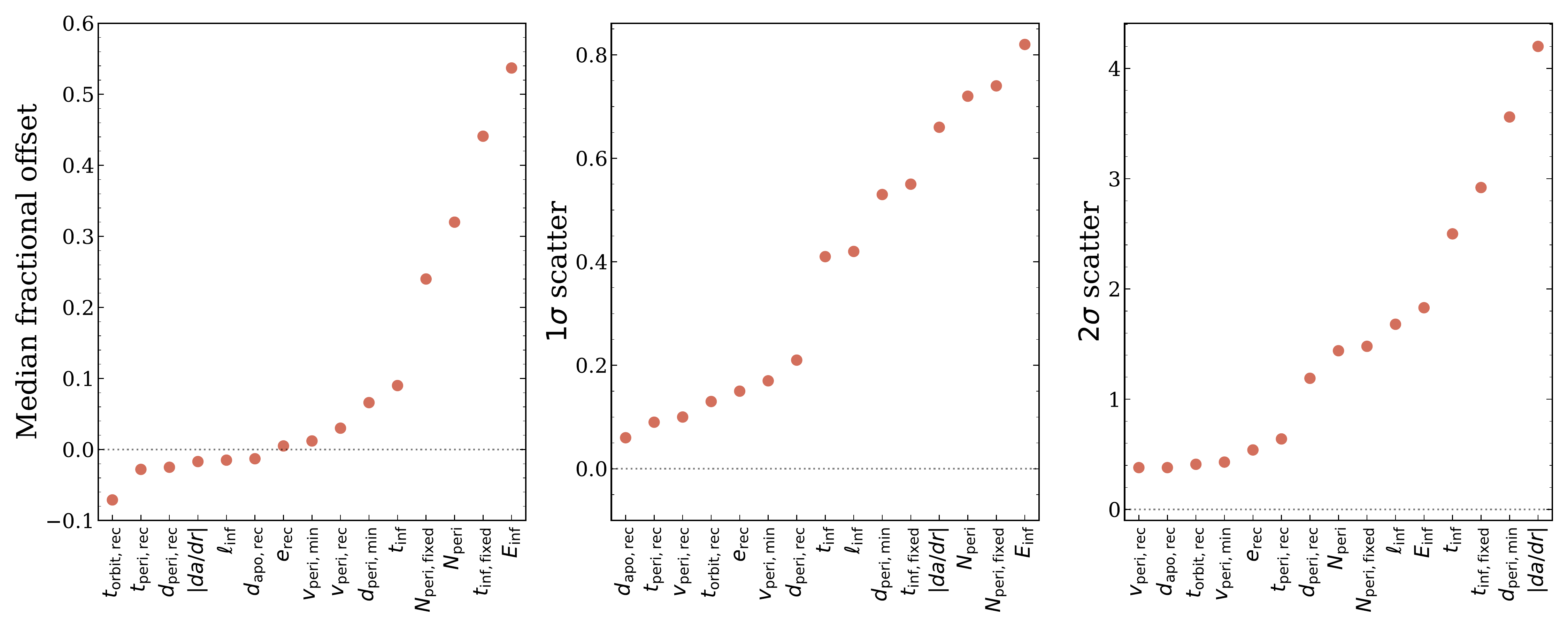}
\end{tabular}
\vspace{-2 mm}
\caption{
Rank ordering the 15 properties of the orbit histories of satellite galaxies, as Table~\ref{tab:summary_stats} lists, based on the fractional level of agreement between orbit modeling in a static axisymmetric host potential and the cosmological baryonic simulations.
For clarity, we shorten $(E_0 - E_{\rm inf}) / U_{\rm 200m,0}$ to $E_{\rm inf}$, and $(\ell_0 - \ell_{\rm inf}) / \ell_0$ to $\ell_{\rm inf}$, and we show the absolute change of $E_{\rm inf}$ in the left panel to more easily compare with the other properties.
The left panel shows the median offset, in order from negative to positive values, and the middle and right panels show the $1-$ and $2 \sigma$ scatters in order of agreement.
Satellite orbit properties that occurred recently, such as the recent pericentre distances/times/velocities, show smaller median offsets, $\lesssim 3$ per cent, compared to properties that occurred further in the past, such as the minimum pericentre distances or infall times, $\gtrsim 5$ per cent.
The same is generally true for the $1 \sigma$ scatter, but not always for the $2 \sigma$ scatter, despite the strong correlation between the two.
Thus, deriving orbit parameters in the \textit{recent} past yields \textit{median} results largely consistent with cosmological simulations, but almost always with significant scatter (uncertainty) always $\gtrsim 10$ per cent, and orbit parameters that occurs further in the past generally suffer from non-trivial bias and \textit{significant} uncertainty.
Furthermore, these results are best-case scenarios for static axisymmetric host potentials, because we fit them near-exact to the simulations at $z = 0$.
}
\label{fig:summary_stats}
\end{figure*}

Figure~\ref{fig:summary_stats} visually represents these summary results, via the median offsets, and $1 \sigma$ and $2 \sigma$ scatters, for the fractional differences between orbit modeling and simulations.
We rank order each property independently in each panel.

The median fractional offset (representing the `bias') of all properties ranges from $-0.54$ for the fractional change in specific energy to $0.44$ for the lookback time of satellite infall using a fixed $\Rthm(\tlb = 0)$.
The properties whose median agrees to within $\pm 5$ per cent across the population include: the most recent pericentre distance, $\dperirec$, the lookback time of the most recent pericentre, $\tperirec$, the maximum value of the derivative of the tidal acceleration, $\dadr$, the fractional change in the angular momentum relative to today, $(\ell_0 - \ell_{\rm infall}) / \ell_0$, the most recent apocentre distance, $\daporec$, the eccentricity of the most recent orbit, $e_{\rm rec}$, and the total satellite velocities at the minimum and most recent pericentres, $\vperimin$ and $\vperirec$, respectively.
Not surprisingly, orbit modeling tends to model/recover \textit{recent} properties of an orbit with the least bias across a population.

Properties that agree moderately, to within $5 - 10$ per cent, include the distance of the minimum pericentre, $\dperimin$, the lookback time of infall into the MW-mass host, $\MWinfall$, and the most recent satellite orbit period, $t_{\rm orbit,rec}$.
Both $\MWinfall$ and $\dperimin$ occurred further back in time than the properties that show the least bias.

Finally, the properties that are most systematically biased in orbit modeling are: the number of pericentric passages both with an evolving and fixed $\Rthm$, $N_{\rm peri}$ and $N_{\rm peri,fixed}$, the lookback time of infall when keeping a fixed $\Rthm = \Rthm(\tlb = 0)$, and the change in orbital energy since infall, $E_{\rm inf}$.


Just as important as examining the bias (offset in the median across the population) is the uncertainty for a given satellite, via the scatter across the population.
This ranges across $\approx 0.06 - 0.82$ at $1 \sigma$ and $\approx 0.38 - 4.2$ at $2 \sigma$.
Again, properties that occurred more recently generally have smaller $1 \sigma$ scatter (aside from $\vperimin$).

At best, the uncertainty for a given satellite is 6 per cent in the apocentre distance, and $\gtrsim 10$ per cent for all other properties.
Additionally, these uncertainties reach nearly a factor of $\approx 2$ in energy, and the $2 \sigma$ scatters are $\gtrsim 40$ per cent.

\textit{Because we model the host potential to within a few per cent at $z = 0$, the uncertainties in Table~\ref{tab:summary_stats} represent lower limits to the bias/uncertainty in orbit modeling in practice.}



\section{Summary \& Discussion} %

\subsection{Summary of results} %
\label{sec:summary}             %

We compared 15 orbit properties for 493 satellite galaxies around 13 MW-mass hosts in the FIRE-2 suite of cosmological baryonic simulations against orbit histories derived from orbit modeling in a static, axisymmetric potential for the same hosts, to quantify rigorously the accuracy and precision of this orbit modeling technique.
Specifically, we fit axisymmetric potentials to each MW-mass hosts at $z = 0$ to within a few percent, which also means that the uncertainties that we present are lower limits to a more realistic scenario applied to the MW/M31 with uncertainty in the underlying host potential.
We now discuss the key questions we raised in the Introduction and our corresponding results. \\

\textit{How much has the mass profile of a MW-mass host evolved over the orbital histories of typical satellites?}
\begin{itemize}
\item Most surviving satellites first fell into their MW-mass halo $3.4 - 9.7 \Gyr$ ago. During that time, $\Mthm$ and $\Rthm$ of the host were $33 - 86$ per cent and $26 - 73$ per cent of their values today (Figure~\ref{fig:hosts}), so they roughly doubled since then.
\item Perhaps more relevant for satellite orbits, the total enclosed mass within a fixed physical distance increased meaningfully (Figures~\ref{fig:mp_vs_lb}-\ref{fig:mp_vs_r}).
Within $50 \kpc$, the typical recent pericentre distance of our satellites, the enclosed mass was only $\approx 74$ per cent of its present-day value at typical satellite infall times ($\approx 7.4 \Gyr$ ago).
\item The fractional increase in the enclosed mass of the host is larger at smaller distances (Figures~\ref{fig:mp_vs_lb}-\ref{fig:mp_vs_r}).
This is contrary to the expectations of `inside-out' growth of a dark-matter halo from DMO simulations, where most halo growth occurs at larger radii \citep[for example][]{Diemand07, Wetzel15_accretion}.
With the inclusion of baryonic physics (most importantly gas cooling), more physical growth occurs at smaller distances, most relevant for satellites with smaller pericentres.
\end{itemize}

\textit{How well does orbit modeling in a static axisymmetric host potential recover key orbital properties in the history of a typical satellite?}
\begin{itemize}
\item Calculating the infall time of a satellite in the model with a growing $\Rthm(t)$ yields more consistent results with the simulations, $\lesssim 2 \Gyr$ offset, compared to using the fixed $\Rthm(t = 0)$ at present-day, but the $1 \sigma$ uncertainties in both metrics can be as high as $\approx 5 - 6\Gyr$ (Figure~\ref{fig:infall_comp}).
\item Orbit history properties that occurred more recently have smaller fractional offsets and uncertainties than properties that occurred in the past.
For instance, the timing and distance of the most recent pericentre have median fractional offsets (uncertainties) $\lesssim 3$ ($9-21$) per cent, compared to the minimum pericentre distance, which occurred further back in time, which has a fractional offset (uncertainty) of $7$ ($53$) per cent (Figures~\ref{fig:infall_comp} and \ref{fig:peri_props} and Table~\ref{tab:summary_stats}).
\item The orbit properties that orbit modeling recovers best (with the smallest bias and uncertainty) include the distance, timing, and velocity of the recent pericentre, the velocity at the minimum pericentre, the most recent apocentre distance, and the orbit eccentricity and period.
The properties that are not recovered well (largest bias and/or uncertainty) include the minimum pericentre distance, number of pericentric passages, the lookback time of infall into MW-mass host halo, maximum strength of the tidal field, and the change in total orbital energy (Figure~\ref{fig:summary_stats} and Table~\ref{tab:summary_stats}).
\item Even with near-perfect knowledge of the mass distribution/potential at $z = 0$ in the host galaxies, the typical uncertainties in these orbit properties range from $6 - 82$ per cent.
Furthermore, the satellite-to-satellite variations in each, that is, the $2 \sigma$ scatters, are $\gtrsim 40$ per cent.
Thus, one cannot recover these orbit properties to within a factor of $\approx 2$ or so, which cautions against overgeneralizing the results for a single satellite from the median trends (Figure~\ref{fig:summary_stats} and Table~\ref{tab:summary_stats}).
\item At fixed mass, the spatial extent or orientation of the host galaxy disc does not significantly affect the orbital properties of the satellites.
Compared to a disc that is rotated by 90 degrees, or a point mass disc model, the median offsets between the fiducial disc model are within $\approx 10^{-3}$ per cent, and the widths of the 68th percentiles are less than $6 \times 10^{-2}$ per cent (Table~\ref{tab:disc_model}).
\end{itemize}

\textit{How far back in time can one reliably model the orbital history of satellites in a static axisymmetric host potential?}
\begin{itemize}
\item The specific energy and specific angular momentum of an orbit are generally not conserved: uncertainties are less than 25 per cent only back to $\approx 3.1 \Gyr$.
Backward integrating orbits more than $\approx 9 \Gyr$ results in energy uncertainties up to a factor of $\approx 2$ or more (Figures~\ref{fig:e_ell_prop}, \ref{fig:conservation} and \ref{fig:summary_stats}, and Table~\ref{tab:summary_stats}).
\item At the most recent orbit, the uncertainty in pericentre distance is already $\sim20$ per cent, and $\sim200\Myr$ in pericentre timing (Figure~\ref{fig:peri_comp}).
Subsequent orbits result in larger uncertainties.
\end{itemize}

\subsection{Discussion} %
\label{sec:disc}        %

\subsubsection{Comparison to D'Souza \& Bell}

Our analysis is closest to that of \citet{DSouza22}, who similarly fit symmetric models to MW-mass halos from the ELVIS suite of DMO simulations \citep{GarrisonKimmel14} to study the uncertainties associated with orbit modeling.
First, we note key differences in methods.
\citet{DSouza22} used DMO simulations, which neglects the (tidal) acceleration from the central galaxy that modify these orbits and could strip/disrupt satellites that orbit nearby.
The internal stellar feedback in a satellite also can reduce the inner density of dark matter and make the satellites more vulnerable to tidal disruption \citep{Bullock17}, although this is a second-order effect \citep{GarrisonKimmel17}.
Without these baryonic effects and processes, the surviving satellites in DMO simulations typically fell into their MW-mass halo earlier and were able to complete more pericentres, while also orbiting \textit{closer} to the center of the host with smaller pericentric passages than we showed in \citet{Santistevan23}.
However, \citet{DSouza22} did account for the effects of dynamical friction in their model, similar to \citet{Patel20}, which acts to slow satellites down and ultimately merge within the host, which we do not.
\citet{DSouza22} also examined the gravitational effects from LMC-like analogs in MW-mass hosts.
Because the LMC is so massive, it hosts its own satellite population, and studies suggest that it is on first infall into the MW and near its first pericentre \citep[for example][]{Kallivayalil13, Deason15, Kallivayalil18, Patel20}, so accounting for its gravitational influence on the surrounding satellites is of great interest.
Although some hosts in our simulations have LMC-like analogues at previous snapshots \citep[see][]{Samuel21, Barry23}, we do not analyze them specifically.

Another significant difference between \citet{DSouza22} and our analysis is that they account for the true mass growth of the MW-mass host at every snapshot by updating their potentials while keeping the potential fixed between snapshots.
We do not account for the mass growth of the MW-mass host, because the majority of orbit modeling studies in the literature implement a fixed mass/potential \citep[for example][]{Patel17, Fritz18, Fillingham19, Pace22}, because we do not know the full mass histories of the MW or M31, and also because the mass assembly history for each MW-mass host in our simulations is unique.

\citet{DSouza22} define a recovered property as being when the absolute value of the fractional difference is less than 30 per cent, that is, $|X_{\rm true} - X_{\rm model}| / X_{\rm true} < 0.3$, and report the fraction of satellites that do not meet this criterion as being `outliers'.
They center their results on two hosts, `iDouglas' which is an isolated MW-mass galaxy and `iOates' which has an LMC analog, but they test orbits in iOates with and without the gravitational contribution from this massive companion.
Similar to our analysis, they focus on the timing and distance of various pericentre events, the apocentre distances, and the infall times of satellites, and find that more recent pericentres and apocentres have smaller outlier fractions than pericentres or apocentres that happened at earlier times.
In particular, the outlier fractions for the most recent, and second-most recent pericentre distances in the hosts without the additional massive satellite are $31.2 - 47$ and $43.8 - 69.9$ per cent, respectively.
Although we do not look specifically at the second-most recent pericentre, we do find that the median fractional offsets and $1 \sigma$ uncertainties for the most recent pericentre distance, $-2.5$ and $21$ per cent, are smaller compared to the same for the minimum pericentre distance which often occurred $\approx 6 \Gyr$ earlier, $0.066$ and $53$ per cent.
The authors also show that the timing of the most recent pericentre is often better recovered than the distance, with an outlier fraction of $13.8-23.2$ per cent.
Our results show that the median fractional offset is $-2.8$, which is comparable to the offset in recent pericentre distance, but with a smaller $1 \sigma$ uncertainty of $9$ per cent.
Thus, it is often easier to recover the timing of a pericentre than its distance.

\citet{DSouza22} similarly showed that the distance of the most recent apocentre has a smaller outlier fraction than the most recent pericentre distance, with a value of only $6.2-34.9$ per cent.
Our work also suggests that the apocentres are easier to recover, with a median fractional offset and $1 \sigma$ uncertainty of $-1.3$ and $6$ per cent, respectively.
They conclude that apocentres are easier to model because they only depend on the binding energy of a satellite galaxy, while the pericentres depend on the angular momentum of a satellite as well as its binding energy.
Properties at apocentre also do not intricately depend on the details of the gravitational potential at small distances like pericentres do, and rather, what is more important is modeling the total enclosed mass precisely, as both studies have done.
Finally, the authors also calculate the infall times of satellites and find good agreement in their simulations and model, with an outlier fraction of $11.2 - 28.9$ per cent.
Although we generally see small median offsets when calculating infall time with an evolving $\Rthm(t)$, $\approx 6.7$ per cent, the associated uncertainty is high, $\approx 41$ per cent, and using a fixed $\Rthm(t = 0)$ is worse.

\citet{DSouza22} explored other models for the growth of the MW-mass host, and most comparable to our work is their results in which they keep the mass fixed over time.
They report that, depending on the property, using the \textit{static} model in some cases produces the best results as compared with the simulations.
However, as one integrates longer in time, the static model becomes less representative of the MW-mass environment, and thus satellite orbit properties that occurred earlier are not modeled as well, which is what we similarly see in the minimum pericentre properties and infall times.
The authors explored a model that accounts for the median mass growth of the 48 MW-mass halos in the ELVIS DMO simulation suite and found that both the static model and median mass growth model reproduce similar results at recent lookback times as well.
Finally, they implemented a static model with 40 per cent more mass, and a static model with different halo concentrations, which both returned larger biases, scatters, and outlier fractions.
Thus, modeling the mass and shape of the halos at present-day is important.

Finally, \citet{DSouza22} compared uncertainties in the virial mass of the MW-mass host, and thus uncertainties in the potential, the uncertainty in the 6D phase-space coordinates of the satellites at $z = 0$, the uncertainty in modeling the recent LMC-like accretion, and the uncertainty in modeling the motion of the MW-mass system as it moves throughout the Universe.
They concluded that the uncertainties in the recovered orbit history properties when using simple parametric forms of the potential or ignoring the LMC-like contribution to the potential are comparable to the uncertainty in results caused by a $\approx 30$ per cent uncertainty in the virial mass of the MW-mass host.

\subsubsection{Comparison to other studies}

The results in Figures~\ref{fig:mp_vs_lb} and \ref{fig:mp_vs_r} (top) differ strikingly from those from DMO simulations, where the enclosed mass at smaller radii is set earlier than at larger radii.
For example, using the Via Lactea dark matter-only simulation, \citet{Diemand07} showed that the enclosed dark matter mass within $100 \kpc$ assembled prior to $z \approx 1.7$ ($\tlb \approx 9.9 \Gyr$ ago), and grew only by about 10 per cent since then.
The enclosed mass at even smaller distances formed earlier.
In a related analysis, \citet{Wetzel15_accretion}, showed that the enclosed dark-matter mass within a fixed physical $r \approx 50 - 100 \kpc$ at $z = 1$ was already $\gtrsim 85 - 95$ per cent of its value at $z = 0$, compared to the mass in stars and gas, which were only $\sim 55 - 70$ per cent of the mass at $z = 0$.
Because dark matter is collisionless and dissipationless, and because the accretion radius grows over time, dark-matter growth occurs largely `inside-out'.
However, gas can cool over time, which drives the formation of the central galaxy, and leads to more physical mass growth at smaller radii compared with the dark matter.

To derive the orbit histories of satellites of the MW and M31, many studies commonly apply orbit modeling in a static host potential or use simple approximations of the growth of the host.
For instance, \citet{Kallivayalil06} used a fixed MW-mass potential, as well as an LMC-like potential, to determine that the SMC was gravitationally bound to the LMC.
\citet{Kallivayalil13} used 3 epochs of HST measurements to constrain the LMC's proper motions and suggest that the LMC is likely on its first infall, as previous studies have suggested \citep[for example][]{Besla07}.
The authors compared different static models of the MW and models that account for the growth of the MW-mass halo over the last $10 \Gyr$, but in the evolving models, the authors did not account for the central galaxy.
\citet{Patel17} similarly sought to understand the orbit of the LMC around the MW, as well as the orbit of M33 around M31, and concluded that like the LMC, M33 is likely fell into the M31 halo less than $\lesssim 2 \Gyr$ ago or so.
Although these authors modeled the many components of the main galaxy, they did not account for any time dependence of the host potential.

\citet{Patel17} also compared the orbits of massive satellite analogues in the dark matter-only Illustris-1-Dark simulation to an NFW model of the MW-mass host halo with a dynamical friction model included.
The authors used the $z=0$ 6D phase-space coordinates for the satellites and integrated their orbits, similar to our pipeline, but for $6\Gyr$, and suggest that this type of orbit modeling technique shows good agreement between the two orbits for satellites on first infall and for satellites that recently completed their first pericentre.
The orbits that we present in Figure~\ref{fig:orbits} show good agreement between the simulations and model for recent lookback time as well, however, only the top-left panel shows good agreement for up to $6\Gyr$.
There are other satellites in our sample that show agreement for this time range, however, we choose to intentionally show test cases in which the model does not do well in this figure as well.

Recent work not only suggests that the LMC has only recently fallen into the MW's halo and is currently near its first pericentre \citep[for example][]{Kallivayalil09, Kallivayalil13}, but that it has a satellite population of its own \citep[for example][]{Deason15, Kallivayalil18, Patel20}.
Other studies account for the gravitational contribution of the LMC on the other satellites of the MW.
\citet{Kallivayalil18} used Gaia data, in conjunction with the DMO Aquarius simulation, to determine which satellites might be gravitationally bound to the LMC.
The simulation does have an LMC analogue, with a similar position and velocity at $z = 0$, but does not account for the gravitational effect of the central galaxy.
\citet{Patel20} used newer Gaia observations and numerically integrated the orbits of satellites in a model of the potential with a MW, LMC, and SMC component to determine the orbital histories of the LMC and its satellites.
Although they keep the potential fixed over time, they only backward integrated the orbits of satellites for $\approx 6 \Gyr$, given that beyond this time, MW-mass halos typically have $\lesssim 80$ per cent of their mass at $z = 0$.
The authors concluded that the derived orbits for satellites of the LMC strongly depend on whether or not you include the LMC in the global potential, and in some cases, the contribution from the SMC is important as well.
Finally, other studies aim to accurately model the LMC potential with basis function expansions fit to simulation data to understand how it affects the MW, and its satellites, as in \citet{GaravitoCamargo19, GaravitoCamargo21, CorreaMagnus22}, but doing so while accounting for the growth of the MW remains challenging.

Studies such as \citet{Fritz18} and \citet{Fillingham19} provide inferences of the infall times, pericentre and apocentre distances, and orbit eccentricities for all satellites in the MW.
\citet{Fritz18} used data from Gaia DR2, and numerically integrated the orbits of all satellites in \textsc{Galpy} in two different models for the MW potential, which they kept fixed over time.
Depending on the mass of the MW, the authors showed that some of the apocentres for the satellites can lie either inside or outside of the virial radius, which has implications for instance in studying how a satellite interacts with the hot gas in the MW halo.
Furthermore, the authors suggested that some satellites have pericentres as close as $\approx 20 \kpc$, where the strong tidal acceleration from the central galaxy is important.
As we showed in Figures~\ref{fig:hosts}-\ref{fig:mp_vs_r}, depending on when these pericentres take place, the mass of the host was only a fraction of its mass today.
\citet{Fillingham19} used the Phat ELVIS DMO simulations, which include an analytic disc potential, to statistically sample satellites with similar present-day positions and velocities to make cosmologically informed predictions for what the infall times of the MW's satellites were.
The Phat ELVIS simulations account for the growth of the disc potential through abundance matching scaling relations, where each disc has a unique growth rate, and thus, tidally disrupt subhalos that orbit close to the disc \citep{Kelley19}.

Most of the  MW-mass simulations in the FIRE-2 suite do not have an LMC-like analogue at $z = 0$, but there are at least four analogues at earlier times, within the last $6 - 7 \Gyr$ ago, and previous works have used these analogues to study both planar configurations of satellites and the effect of LMC-mass satellites on subhalo populations \citep{Samuel21, Barry23}.
Given that the LMC recently fell into the MW's dark matter halo, $\lesssim 2 \Gyr$ ago \citep[for example][]{Besla07, Kallivayalil13, Patel17}, previous orbit modeling studies that include the effects of the LMC to derive orbits of the MW's satellites work well in this recent regime.
On the other hand, because we do not include the effects of an LMC in the host potential, our work informs how well static potential orbit modeling works in the regime before the LMC was a satellite, that is, $\gtrsim 2 \Gyr$ ago.
However, the results suggesting that the LMC is on first infall and just passed its first pericentre are trustworthy, given that the mass of the MW and its dark matter halo have not significantly changed over the last $\lesssim2\Gyr$.
As such, both kinds of studies, both with and without an LMC analogue, complement each other in understanding the full, complex MW formation history.


\section*{Acknowledgements}

We gratefully appreciate discussions with, and comments from, Jo Bovy, Richard D'Souza, Chris Fassnacht, Nicol\'{a}s Garavito-Camargo, and Gurtina Besla.

We used NumPy \citep{NumPy}, SciPy \citep{SciPy}, AstroPy \citep{AstroPy}, and Matplotlib \citep{Matplotlib}.
We also extensively used GalPy \citep{Bovy15}, where the user guide and github repository are available at  \url{https://docs.galpy.org} and \url{https://github.com/jobovy/galpy}, respectively.
To creatively generate unique color hex codes for some of the figures, we also supplied prompts to the AI drawing model \texttt{craiyon}\footnote{\url{https://www.craiyon.com}} (formerly, \textsc{DALL$\cdot$E} mini) to create distinct color palettes.
We adopted additional color schemes from the palettes provided by Pantone from the recent JWST press release images.

IBS received support from NASA, through FINESST grant 80NSSC21K1845.
AW received support from: NSF via CAREER award AST-2045928 and grant AST-2107772; NASA ATP grant 80NSSC20K0513; HST grants AR-15809, GO-15902, GO-16273 from STScI.
JM is funded by the Hirsch Foundation.
We ran simulations using: XSEDE, supported by NSF grant ACI-1548562; Blue Waters, supported by the NSF; Frontera allocations AST21010 and AST20016, supported by the NSF and TACC; Pleiades, via the NASA HEC program through the NAS Division at Ames Research Center.


\section*{Data Availability} %

The data in the figures above, and the python code that we used to analyze these data, are available at \url{https://ibsantistevan.wixsite.com/mysite/publications}, which uses the publicly available packages \url{https://bitbucket.org/awetzel/gizmo\_analysis}, \url{https://bitbucket.org/awetzel/halo\_analysis}, and \url{https://bitbucket.org/awetzel/utilities}.
The FIRE-2 simulations are publicly available \citep{Wetzel23} at \url{http://flathub.flatironinstitute.org/fire} which include the star, gas, and dark matter particle data at various redshifts from $z = 0 - 10$, along with the halo merger trees, which include the positions and velocities of all satellites at all 600 snapshots.
Additional FIRE simulation data is available at \url{https://fire.northwestern.edu/data}.
A public version of the GIZMO code is available at \url{http://www.tapir.caltech.edu/~phopkins/Site/GIZMO.html}.




\bibliographystyle{mnras}
\bibliography{orbit_modeling_ii}


\appendix

\section{Modeling host mass profiles} %
\label{app:mass_profile}                       %

We integrate the orbits of satellite galaxies using the galactic dynamics Python package \galpy \citep{Bovy15}, which allows users to define custom host potentials.
We thus fit the enclosed mass profiles of the MW-mass hosts in our simulations as a sum of disc and halo components, to supply to \galpy.

\subsection{Modeling the disc} %
\label{sec:disc_model}         %

We model the MW-mass disc as the sum of two double exponential discs, one for the inner disc/bulge, and one for the outer disc.
The density profile of the disc is:
\begin{equation}
\rho(R,Z) = A^{\rm inner}_{\rm disc} \ e^{-R / R_{\rm disc}^{\rm inner} - |Z| / h_z}\ + A^{\rm outer}_{\rm disc} \ e^{-R / R_{\rm disc}^{\rm outer} - |Z| / h_z}
\label{eq:disc_den}
\end{equation}
where $A^{\rm inner}_{\rm disc}$ and $A^{\rm outer}_{\rm disc}$ are amplitudes of mass density ($M_\odot / \kpc^3$), $R_{\rm disc}^{\rm inner}$ and $R_{\rm disc}^{\rm outer}$ are disc scale radii ($\kpc$), and $h_z$ is the disc scale height ($\kpc$).
Given that satellite orbits are more sensitive to the enclosed \textit{mass}, rather than the local density, we first integrate out the vertical component of Equation~\ref{eq:disc_den}, then we integrate over cylindrical $R$ to obtain the disc enclosed mass given by
\begin{equation}
M(< R) = 4 \pi A h_z R_{\rm inner} \bigg[ R_{\rm inner} - e^{-R / R_{\rm inner}}(R_{\rm inner} + R) \bigg],
\label{eq:disc_mass}
\end{equation}
and a similar term for the outer disc component.
We thus ensure that the total enclosed mass agrees to within a few percent at all relevant radii when integrated over all space.
We fit and model the disc with star particles and cold gas ($T < 10^5 \K$) within $R = 0 - 20 \kpc$ and $|Z| < 3 \kpc$.
Including mass within $|Z| < 4, 5$ did not significantly change the derived parameters.

The orbits of typical satellites are not sensitive to the details of the size, geometry, or orientation of the disc, see Appendix~\ref{app:point_mass}.

\subsection{Modeling the halo}
\label{sec:halo_model}

To fit the halo component, we use a generalized form of the spherical Navarro-Frenk-White \citep[NFW,][]{NFW} density profile:
\begin{equation}
\rho(r) = \frac{A_{\rm halo}}{4 \pi a_{\rm halo}^3}\frac{1}{(r/a_{\rm halo})^{\alpha}(1 + r / a_{\rm halo})^{\beta - \alpha}}
\label{eq:halo_den}
\end{equation}
where $A_{\rm halo}$ is the amplitude ($\Msun$), $a_{\rm halo}$ is the scale radius ($\kpc$), and $\alpha$ and $\beta$ are the slopes of the inner and outer density profile.
We then integrate the analytic form of $\rho(r)$ to convert this to an enclosed mass profile:
\begin{equation}
M(< r) = \frac{A_{\rm halo}}{3 - \alpha} \bigg(\frac{r}{a}\bigg)^{3 - \alpha} {}_2F_{1}(3 - \alpha, - \alpha + \beta; 4 - \alpha; -r / a_{\rm halo})
\label{eq:halo_mass}
\end{equation}
where ${}_2F_1$ is the Gauss hypergeometric function.
We model and fit this profile with dark-matter and hot gas ($T > 10^5 \K$) within $r < 10 \kpc$, and \textit{all} particles (dark matter, all gas, and stars) at $r = 10 - 500 \kpc$.
We obtained similar parameters selecting particles within $r < 300, 350, 400 \kpc$, but we use the fits out to 500 kpc, because many satellites orbit out to there \citep[see][]{Santistevan23}.
We tried fitting the halos to a regular NFW profile, where $\alpha = 1$ and $\beta = 3$, but we obtained notably better agreement with the generalized form above.

\subsection{Total mass profile}
\label{sec:full_model}

We use the parameters we derived from fitting the above analytic density/mass profiles in their respective potentials in \galpy.
In \galpy, we specifically use the spherically symmetric `TwoPowerSphericalPotential' for the DM halo, and axisymmetric `DoubleExponentialdiscPotential' for the inner and outer discs, and we input all three to define the total potential of a given host, to use to integrate the orbits of its satellites.
Table~\ref{tab:param} lists the fit parameters.

\begin{table*}
\centering
\caption{
Best-fit parameters to the double-exponential disc profile (Equations~\ref{eq:disc_den}-\ref{eq:disc_mass}) and the generalized NFW profile for the halo (Equations~\ref{eq:halo_den}-\ref{eq:halo_mass}).
Columns: Name of host; halo amplitude, $A_{\rm halo}$; halo scale radius, $a_{\rm halo}$; halo inner slope, $\alpha$; halo outer slope, $\beta$; disc inner amplitude, $A^{\rm inner}_{\rm disc}$; disc inner scale length, $R_{\rm inner}$; disc outer amplitude, $A^{\rm outer}_{\rm disc}$; disc outer scale length, $R_{\rm outer}$; disc scale height, $h_z$.
}
\begin{tabular}{|c|c|c|c|c|c|c|c|c|c|}
\hline
\hline
Name & $A_{\rm halo}$ & $a_{\rm halo}$ & $\alpha$ & $\beta$ & $A_{\rm disc}^{\rm inner}$ & $R_{\rm disc}^{\rm inner}$ & $A_{\rm disc}^{\rm outer}$ & $R_{\rm disc}^{\rm outer}$ & $h_{\rm z}$ \\
& [$10^{11} \Msun$] & [$\kpc$] & & & [$10^{9} \Msun \kpc^{-3}$] & [$\kpc$] & [$10^{8} \Msun \kpc^{-3}$] & [$\kpc$] & [$\kpc$] \\
\hline
m12m    & 3.78 & 17.32 & 1.57 & 2.78 & 6.47 & 0.79 & 7.98 & 4.41 & 0.64 \\
Romulus & 5.30 & 6.65  & 0.00 & 2.87 & 8.31 & 0.86 & 2.10 & 8.04 & 0.55 \\
m12b    & 3.84 & 16.89 & 1.48 & 2.82 & 18.0 & 0.65 & 6.08 & 4.24 & 0.51 \\
m12f    & 3.65 & 14.77 & 1.45 & 2.74 & 8.39 & 0.84 & 2.11 & 7.40 & 0.54 \\
Thelma  & 4.49 & 8.84  & 0.40 & 2.90 & 3.30 & 1.04 & 1.98 & 6.28 & 0.75 \\
Romeo   & 8.99 & 29.91 & 1.47 & 3.23 & 4.61 & 1.02 & 2.25 & 6.64 & 0.55 \\
m12i    & 5.70 & 32.15 & 1.58 & 2.99 & 9.10 & 0.78 & 3.87 & 4.56 & 0.55 \\
m12c    & 7.93 & 24.37 & 0.90 & 3.05 & 7.05 & 0.71 & 6.41 & 3.65 & 0.57 \\
m12w    & 2.99 & 22.87 & 1.61 & 2.73 & 5.53 & 0.68 & 11.7 & 2.19 & 0.67 \\
Remus   & 8.32 & 36.49 & 1.56 & 3.21 & 4.68 & 0.90 & 1.60 & 6.50 & 0.54 \\
Juliet  & 2.85 & 9.57  & 0.87 & 2.82 & 6.97 & 0.77 & 1.16 & 6.32 & 0.55 \\
Louise  & 2.30 & 6.41  & 0.29 & 2.76 & 1.92 & 0.99 & 0.54 & 9.36 & 0.58 \\
m12z    & 3.22 & 33.96 & 1.58 & 2.69 & 0.59 & 0.40 & 1.73 & 2.94 & 1.54 \\
\hline
\hline
\end{tabular}
\label{tab:param}
\end{table*}

\begin{figure}
\centering
\begin{tabular}{c}
\includegraphics[width = 0.95 \linewidth]{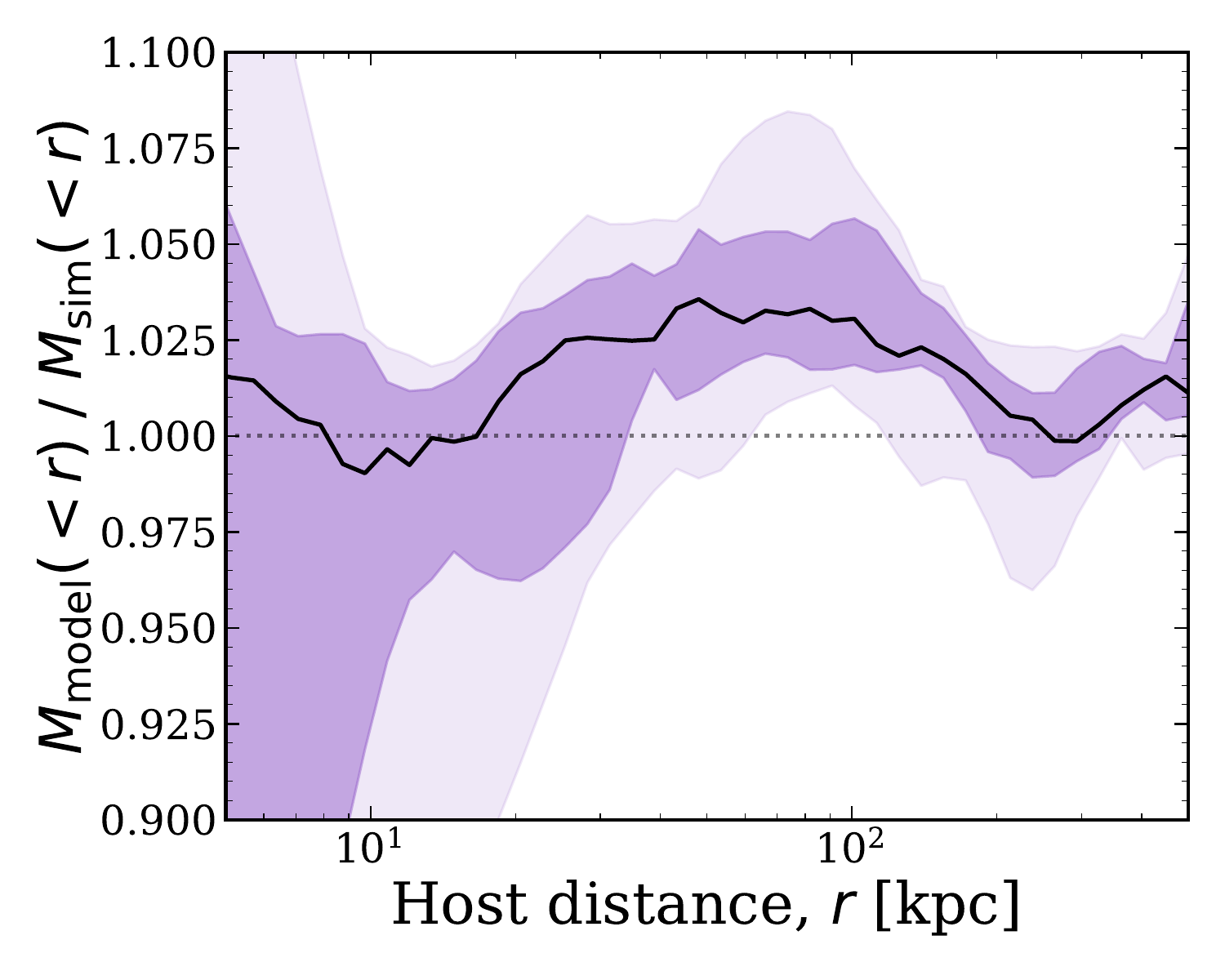}
\end{tabular}
\vspace{-2 mm}
\caption{
Ratio of the enclosed total mass in each best-fit model to that in each simulation, as a function of distance from the center of the MW-mass host, $r$, at $z = 0$.
We fit a double-exponential profile for the inner and outer disc (Equation~\ref{eq:disc_den}-\ref{eq:disc_mass}) and a generalized NFW profile for the halo (Equations~\ref{eq:halo_den}-\ref{eq:halo_mass}).
Table~\ref{tab:param} lists all fit parameters.
Black line shows the median and the shaded regions show the 68th and 95th percentile scatter across our 13 hosts.
The median agrees to within $< 4$ per cent at all radii and $\lesssim 1$ per cent for the mass within $\Rthm$ ($344 - 472 \kpc$).
The ratio shows larger host-to-host scatter at $\lesssim 10 \kpc$, but only $\approx 5$ per cent of satellites in our sample orbit this close.
This level of agreement ensures that modeling the total (axisymmetric) mass profile of each host at $z = 0$ is not a significant source of error for our orbit modeling.
}
\label{fig:mp_fit}
\end{figure}

Figure~\ref{fig:mp_fit} shows the ratio of the halo+disc analytic model using the parameters in Table~\ref{tab:param} to the enclosed masses in the simulations; a 1:1 ratio is presented as the dotted horizontal line.
At all distances we show, we model the median enclosed mass to within 3-4 per cent, and the 68th percentile range is within 5 per cent beyond $10 \kpc$.
Within $10 \kpc$, the 68th and 95th percentiles of the enclosed mass ratios span both higher and lower than what we present in the y-axis, but only $\approx 5$ per cent of surviving satellites orbited at such small distances.

Individually, we model the two-component disc to within $\approx 4$ per cent at $R = 2 - 20 \kpc$ and the halo to within $\approx 10 - 20$ per cent.
Although we double counts the star particles and cold gas between $10 - 20 \kpc$, the median mass ratio in Figure~\ref{fig:mp_fit} is $\lesssim 2$ per cent within this distance and produces a good fit to the data.
One galaxy, m12c, experienced a significant merger ($\approx 3:1$ mass ratio) $\approx 9 \Gyr$ ago, thus modeling this host with symmetric analytic models does not capture the complexity in its mass distribution.
Thus, the total enclosed mass in m12c is fit to within $\approx 15$ per cent.

\section{Alternative models for the host galaxy potential: rotated disc and point mass}
\label{app:point_mass}

Here, we investigate the results of using different potential models for the central galaxy, including (1) a disc that is rotated by 90 degrees, and (2) a point mass.
We compare these models with the fiducial disc model in Table~\ref{tab:param}.
We numerically integrate the satellites in these alternative models and calculate the same orbit properties in Table~\ref{tab:summary_stats}.
For each property $X$, we present only the fractional differences between the models, $(X_{\rm disc, rot} - X_{\rm disc}) / X_{\rm disc}$ and $(X_{\rm point} - X_{\rm disc}) / X_{\rm disc}$, where $X_{\rm disc, rot}$ refers to the fiducial disc model rotated by 90 degrees, $X_{\rm point}$ refers to the point mass model, and $X_{\rm disc}$ is our fiducial model.
Table~\ref{tab:disc_model} shows the median offset and width of the 68th percentile.
For cases in which the median is zero, we instead show the mean offset and standard deviation.

\begin{table*}
\caption{
Similar to Table~\ref{tab:param}, but now comparing our fiducial model for the host disc to models in which (1) we rotate the disc by 90 degrees or (2) use a simple point mass.
We show only the fractional differences for each property $X$ via $(X_{\rm disc,rot} - X_{\rm disc}) / X_{\rm disc}$ for the rotated disc model (middle columns), and $(X_{\rm point} - X_{\rm disc}) / X_{\rm disc}$ for the point mass model (right columns).
Column list: property name, variable, median offset, $1 \sigma$ scatter.
We do not show the minimum pericentre properties given that the orbits in the models are periodic.
Values with an asterisk at the end represent cases where we present the mean offset or standard deviation, instead of median and $1 \sigma$ scatter, because they are discrete quantities or have medians of 0.
The overall bias in any orbit property for both alternative models is less than 1 per cent, and the $1 \sigma$ uncertainties only reach as large as $\approx 7$ per cent for the number of pericentric passages.
\textit{Thus, the orientation and geometry of the central disc are not important factors in orbit modeling uncertainties.}
}
\begin{tabular}{|c c |@{\vline}| c c |@{\vline}| c c|}
\hline
\hline
& & \multicolumn{2}{c}{\underline{Rotated disc}} & \multicolumn{2}{c}{\underline{Point Mass}} \\
Property & Variable & Median offset  & $1 \sigma$ scatter  & Median offset & $1 \sigma$ scatter  \\
\hline
Recent pericentre distance                           & $d_{\rm peri, rec}$   & $2\times10^{-5}$   & $2\times10^{-3}$    & $5\times10^{-3}$   & $7\times10^{-3}$  \\
Timing of recent pericentre                          & $t_{\rm peri, rec}$   & $1\times10^{-9}$   & $1\times10^{-3}$    & $3\times10^{-3}$   & $7\times10^{-3}$  \\
Number of pericentric passages                       & $N_{\rm peri}$        & $2\times10^{-3*}$  & $5\times10^{-2*}$   & $-6\times10^{-3*}$ & $4\times10^{-2*}$ \\
Pericentric passages post-infall at $R_{\rm 200m,0}$ & $N_{\rm peri, fixed}$ & $2\times10^{-3*}$  & $5\times10^{-2*}$   & $-8\times10^{-3*}$ & $7\times10^{-2*}$ \\
Velocity at recent pericentre                        & $v_{\rm peri, rec}$   & $-3\times10^{-5}$  & $2\times10^{-3}$    & $-5\times10^{-3}$  & $7\times10^{-3}$  \\
Recent apocentre distance                            & $d_{\rm apo, rec}$    & $4\times10^{-7}$   & $4\times10^{-4}$    & $1\times10^{-3}$   & $4\times10^{-3}$  \\
Lookback time of infall                              & $t_{\rm inf}$         & $-2\times10^{-3*}$ & $1.5\times10^{-3}$  & $3\times10^{-3}$   & $5\times10^{-3}$  \\
Lookback time of infall at $R_{\rm 200m,0}$          & $t_{\rm inf, fixed}$  & $-1\times10^{-7}$  & $2.5\times10^{-3}$  & $-2\times10^{-8}$  & $3\times10^{-3}$  \\
Recent orbit eccentricity                            & $e_{\rm rec}$         & $-4\times10^{-5}$  & $2\times10^{-3}$    & $-1\times10^{-3}$  & $5\times10^{-3}$  \\
Recent orbit period                                  & $T_{\rm rec}$         & $3\times10^{-8}$   & $2\times10^{-3}$    & $7\times10^{-3}$   & $1\times10^{-2}$  \\
Max strength of tidal force                          & $\dadr$               & $1\times10^{-3*}$  & $3\times10^{-2*}$   & $-2\times10^{-3}$  & $2\times10^{-2}$  \\
\hline
\hline
\end{tabular}
\label{tab:disc_model}
\end{table*}

For the rotated disc model, the median fractional difference across all orbit properties is $\approx 10^{-3}$ or smaller, and the $1 \sigma$ scatters are less than 5 per cent.
We also calculate the median fractional difference only for satellites that orbit closest to the disc, with pericentres $< 50 \kpc$, and the offsets are still $\lesssim 10^{-3}$ per cent.
\textit{Thus, we conclude that the orientation of the host galaxy isk is not important for orbit modeling.}

To implement a point mass potential, we use \galpy's `KeplerPotential', which takes only the mass as the parameter.
We integrate the enclosed galaxy mass profile in Equation~\ref{eq:disc_mass} out to $30 \kpc$ and input this into the point mass potential.
Table~\ref{tab:disc_model} shows that, even for this simplest-possible model for the central galaxy mass distribution, the median offsets and percentile widths are $\lesssim 7$ per cent.
\textit{This reinforces that the details of the mass distribution of the central galaxy are not important; what matters only is modeling the total baryonic mass of the central galaxy.}

We also tried `shrinking' our fiducial disc model, by reducing the scale radii and scale height to 90, 50, 10, and 1 per cent of their best-fit values in Table~\ref{tab:param}, which showed results intermediate between the rotated disc and point mass models.

These results are consistent with the tests of \citet{GarrisonKimmel17}, who compared the surviving subhalo populations in two of the FIRE-2 MW-mass hosts against DMO simulations with an embedded disc potential.
They showed that the number of surviving subhalos in the simulations with an embedded disc potential agrees well with those in the baryonic FIRE-2 simulations, both of which are much smaller than in a DMO simulation without an embedded disc.
They additionally tested various embedded disc potentials by doubling the scale length, fixing the scale height to 1 pc, doubling the total disc mass, including the gas mass to the disc, and implementing a Hernquist sphere instead of their Miyamoto-Nagai disc potential.
Most of these alternate disc potentials led to similar results, reinforcing that the details of the shape of the central galaxy potential are less important than simply its overall mass.

\section{Comparing discrete pericentres} %
\label{app:peri_comp}                    %

\begin{figure}
\centering
\begin{tabular}{c}
\includegraphics[width = 0.95 \linewidth]{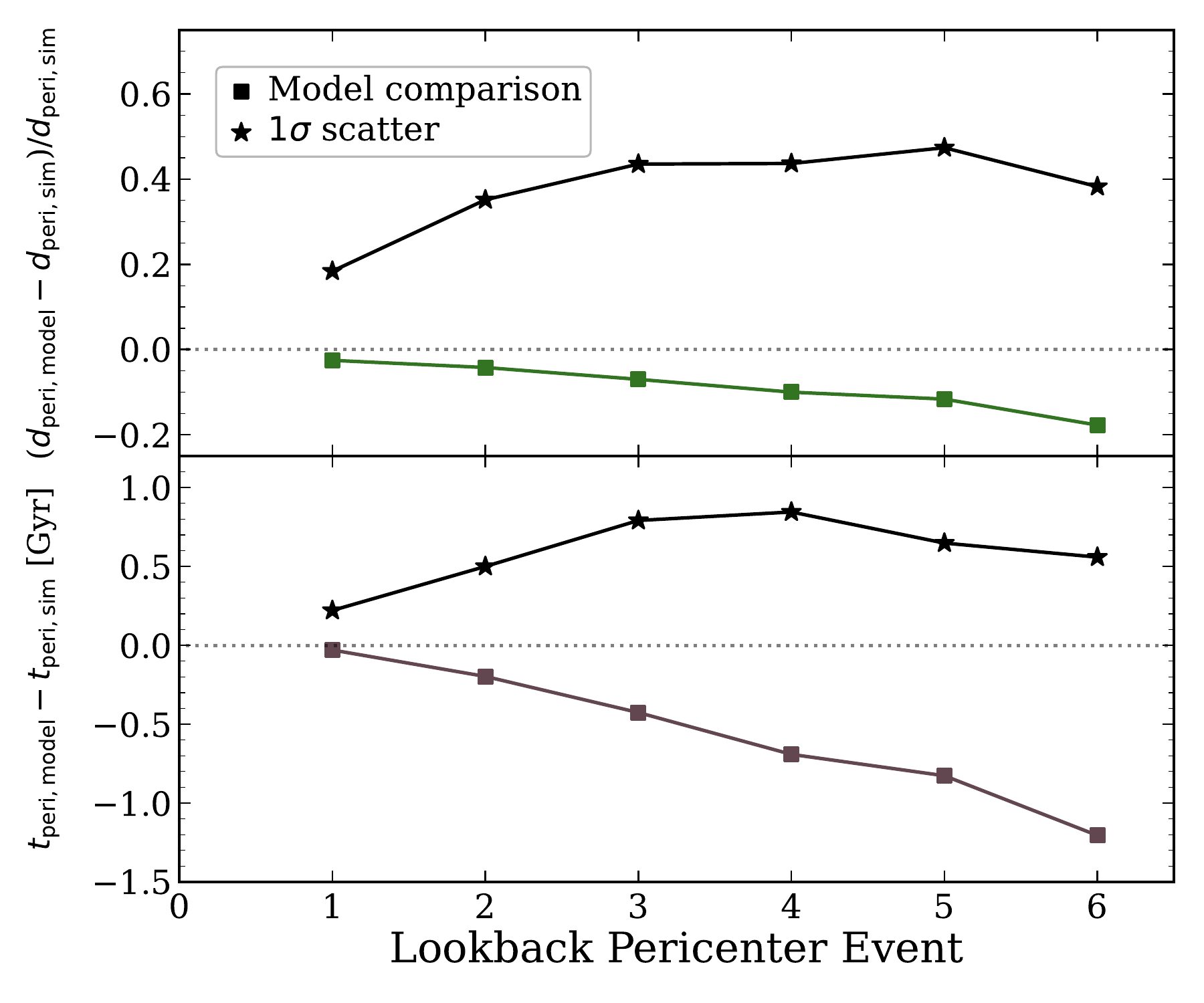}
\end{tabular}
\vspace{-2 mm}
\caption{
Comparing the timing and distance of each pericentric passage between orbit modeling and simulations, which we measure at each `pericentre event', where 1 is the most recent pericentre, 2 is the second most recent, and so forth.
We select all satellites that experienced (at least) a given number of pericentres, and we show the median and $1 \sigma$ scatter across all satellites.
\textbf{Top}: The median fractional difference in pericentre distance decreases from $-2.5$ to $-18$ per cent going back 6 pericentres while the $1 \sigma$ scatter increases from $18$ to $48$ per cent.
Less than $3$ per cent of satellites experienced more than 7 pericentres, in both the simulations and orbit models.
\textbf{Bottom}: The median difference in the timing of the pericentric events decreases with lookback pericentre events from $-30 \Myr$ for the most recent pericentre to $-1.2 \Gyr$.
The $1 \sigma$ scatter ranges from $\approx 0.2 - 0.85 \Gyr$, so not only does orbit modeling increasingly systematically underpredict the timing of a pericentre, but also, orbit modeling becomes more uncertain, over longer lookback times.
}
\label{fig:peri_comp}
\end{figure}

To understand better how well orbit modeling and cosmological simulation agree \textit{at each orbit}, Figure~\ref{fig:peri_comp} compares the timing and distance of each pericentric event.
We select all satellites at a given `lookback pericentre event', where 1 is the most recent pericentre, 2 is the second-most recent, and so forth.
We plot the fractional and raw differences in the pericentre distance and timing.

In our sample, $86$ per cent of satellites experienced at least one pericentre, $\approx 47$ per cent experienced two or more, and $\approx 30$ per cent experienced three or more.
The most pericentres experienced is 10, but we cut the figure at 6 pericentres given that only $3.4$ per cent experienced more than this.

Focusing on trends with distance, the median fractional difference in the most recent pericentre is only $-2.5$ per cent, similar to Figure~\ref{fig:peri_props} (top row).
As we compare pericentres further back in time, the median decreases to $-18$ per cent for satellites at their 6th-most recent pericentre.
However, this is not to say that the model \textit{never} overpredicts the pericentre distances.
For the most recent pericentric passage, the model under-predicts the distance in $\approx 60$ per cent of the sample, and over-predicts the distance in the other 40 per cent.

Likewise, the $1 \sigma$ scatter generally increases with increasing lookback pericentre events, where it reaches a max value of $\approx 48$ per cent at the 5th-most recent pericentre.
Thus, the median difference between the model and simulation increases with each orbit as we look back in time, but also the uncertainties associated with these pericentre distances increases.

Similar to trends in distance, Figure~\ref{fig:peri_comp} (bottom) shows that the median difference in the timing of each pericentre decreases from $-30$ Myr at the most recent pericentre to $-1.2 \Gyr$ for the 6th most recent.
Orbit modeling continuously under-predicts pericentre events, presumably because the host mass is unchanging as the satellites orbit.
For a given satellite, the more massive host with a deeper gravitational potential will result in a more bound orbit compared to the same satellite in the same MW-mass host with less mass at earlier times.
More bound satellites thus orbit deeper in the potential with shorter orbit timescales.
Regarding the most recent pericentric passage, the model under-predicts the timing in $\approx 66$ per cent of satellites and over-predicts the timing in $30$ per cent; the remaining 4 per cent of satellites have nearly identical values.
The $1 \sigma$ scatter spans $\gtrsim 0.5 \Gyr$ for the second-most recent pericentre and beyond.

These results highlight another aspect of how far back in time orbit modeling reliably works.
If one is only interested in deriving the most recent pericentre distance or time, the model recovers the median trends of satellites to within a few per cent, but the uncertainties are non-negligible.
For the satellites that orbited more than once, the most recent pericentre is not always the smallest \citep[see][]{Santistevan23}, therefore if one is interested in how close satellites orbited to their host, then the uncertainty in deriving these distances will be $\gtrsim 35$ per cent, and $\gtrsim 0.5 \Gyr$ in the timing.

\bsp	
\label{lastpage}
\end{document}